\documentclass[showpacs,aps,prc,preprint,nofootinbib,superscriptaddress]{revtex4}

\usepackage[utf8]{inputenc}
\usepackage{graphicx}
\usepackage{subfigure}
\usepackage{float}
\usepackage{amsmath}
\usepackage{amssymb}
\usepackage{color}
\usepackage{slashed}
\usepackage{morefloats}

\def\bm{\boldsymbol}
\newcommand{\beq}{\begin{equation}}
\newcommand{\eeq}{\end{equation}}
\newcommand{\bea}{\begin{eqnarray}}
\newcommand{\eea}{\end{eqnarray}}
\newcommand{\be}{\begin{eqnarray}}
\newcommand{\ee}{\end{eqnarray}}
\newcommand{\no}{\nonumber \\}

\def\vp{{\bm p}}
\def\vq{{\bm q}}
\def\vk{{\bm k}}

\def\vx{{\bm x}}
\def\vy{{\bm y}}

\def\vr{{\bm r}}
\def\vs{{\bm\sigma}}

\newcommand{\simge}{\hspace*{0.2em}\raisebox{0.5ex}{$>$}
     \hspace{-0.8em}\raisebox{-0.3em}{$\sim$}\hspace*{0.2em}}
\newcommand{\simle}{\hspace*{0.2em}\raisebox{0.5ex}{$<$}
     \hspace{-0.8em}\raisebox{-0.3em}{$\sim$}\hspace*{0.2em}}

\begin{document}

\title{Triton and Neutron-Deuteron Scattering\\
up to Next-to-Leading Order in Chiral EFT}

\author{Young-Ho Song}
\email[]{yhsong@ibs.re.kr}
\affiliation{Rare Isotope Science Project, Institute for Basic Science, 
Daejeon 305-811, Korea}

\author{Rimantas Lazauskas}
\email[]{rimantas.lazauskas@iphc.cnrs.fr}
\affiliation{Universit\'e de Strasbourg, CNRS IPHC UMR 7178,
F-67000 Strasbourg, France}

\author{U. van Kolck}
\email[]{vankolck@ipno.in2p3.fr}
\affiliation{Institut de Physique Nucl\'eaire,
CNRS/IN2P3, Univ.~Paris-Sud, Universit\'e Paris-Saclay, F-91406 Orsay Cedex,
France}
\affiliation{Department of Physics, University of Arizona,
Tucson, AZ 85721, USA}


\begin{abstract}
Determination of the proper power-counting scheme
is an important issue for the systematic
application of Chiral Effective Field Theory in nuclear physics.
We analyze the
cutoff dependence of three-nucleon observables (the
neutron-deuteron scattering lengths and the triton binding energy)
at the leading
and next-to-leading orders
of a power counting that ensures order-by-order renormalization
in the two-nucleon system.
Our results
confirm that, as usually assumed in the literature,
three-body forces are not needed for renormalization
of the three-nucleon system up to next-to-leading order.
\end{abstract}


\maketitle

\section{Introduction \label{sec:Int}}

Hadronic effective field theories (EFTs) provide a representation
of QCD at the relatively large distances involved in nuclear
dynamics. 
Chiral EFT, which includes nucleons, pions and the
lightest baryon excited states as active degrees of freedom,
aims at an expansion of
nuclear amplitudes in powers of $Q/M_{QCD}$, where the typical external
momentum is of the order of the pion mass, $Q\sim m_\pi$,
and $M_{QCD}\sim 1$ GeV is the characteristic scale of QCD.

The investigation of nuclear forces and currents
in Chiral EFT was initiated in the early 1990s
\cite{Weinberg:1990rz,Rho:1990cf,Weinberg:1991um,Ordonez:1992xp,Weinberg:1992yk,Ordonez:1993tn,vanKolck:1994yi,Ordonez:1995rz},
based on a two-step scheme proposed by Weinberg.
The first step is to calculate the nuclear
potential and currents --- the sum of ``irreducible'' diagrams ---
up to a given order, that is, a given power of $Q/M_{QCD}$. The
contributions of each order are given by the usual power counting
of ChPT, assuming that short-range nuclear 
interactions obey the same naive dimensional analysis (NDA)
\cite{Manohar:1983md} used in the purely perturbative problems
involving at most one nucleon. The second step is to solve the
Schr\"odinger or Lippmann-Schwinger (LS) equations exactly with
the potential truncated at the desired order, which corresponds to
the non-perturbative iteration of the potential subdiagrams.
Following the initial successes in explaining some of the
qualitative features of models
\cite{Rho:1990cf,Weinberg:1991um,Ordonez:1992xp,Weinberg:1992yk,Ordonez:1993tn,vanKolck:1994yi,Ordonez:1995rz},
the application of ChPT to nuclear forces 
and currents has reached the point
where a very accurate description of data can be achieved
--- see, for example,
Refs. \cite{Bedaque:2002mn,Epelbaum:2008ga,Machleidt:2011zz}
and references therein.

Renormalization-group invariance (RGI) is the crucial ingredient that separates 
nuclear EFTs from the phenomenological models that preceded them.
RGI guarantees control over arbitrary choices made during the
regularization procedure, such as the functional form of the
regulator and the numerical value of the regulator parameter(s). 
Although initial numerical results using Gaussian regulators
with a cutoff in the range $\Lambda=500\to 1000$ MeV seemed to
indicate only moderate cutoff dependence \cite{Ordonez:1995rz},
examples have accumulated since the mid-90s showing that
Weinberg's scheme is not consistent with RGI in the two-nucleon
sector. 

In the spin-singlet $S$ wave ($^1S_0$), NDA prescribes a
single chiral-invariant counterterm at leading order (LO) together
with one-pion exchange (OPE), but RGI demands also a
chiral-breaking counterterm \cite{Kaplan:1996xu,Beane:2001bc},
which according to NDA would appear only at relative 
${\cal O}(Q^2/M_{QCD}^2)$, or next-to-next-to-leading order (N$^2$LO)
\footnote{Note that in part of the literature
\cite{Epelbaum:2008ga,Machleidt:2011zz} this order is denoted as
next-to-leading order (NLO) because {\it under the assumption of
NDA} parity and time reversal ensure that ${\cal O}(Q/M_{QCD})$
contributions vanish. We prefer to denote ${\cal O}(Q^n/M_{QCD}^n)$ 
corrections as N$^n$LO to accommodate the known violation of NDA.}.
Contrary to the LO chiral-invariant counterterm,
this chiral-breaking interaction involves pions.
Its enhancement over NDA affects processes with
external pions, but has only mild effects on the two-nucleon
amplitude at a fixed pion mass. However, other two-nucleon channels suffer
from larger cutoff effects stemming from the singular nature of
the OPE tensor force. In each spin-triplet channel, where the OPE
tensor force is attractive, the solution of the LS equation
generates bound states that cross threshold as the cutoff
$\Lambda$ increases beyond about 1 GeV
\cite{Nogga:2005hy,PavonValderrama:2005uj}. In Weinberg's scheme
OPE is iterated in all waves, but NDA prescribes an LO counterterm
only in the $^3S_1$ wave. NDA erroneously assigns counterterms in
the attractive $P$, $D$, ... waves to N$^2$LO, N$^4$LO, ...,
respectively. At higher orders in Weinberg's scheme, cutoff
dependence does not disappear
\cite{PavonValderrama:2005wv,Entem:2007jg,Yang:2009kx,Yang:2009pn,Zeoli:2012bi}.

One way to achieve RGI is to
relegate pion exchange 
to higher orders, which are treated in
perturbation theory at the amplitude level
\cite{Kaplan:1998tg,Kaplan:1998we}. Unfortunately,
in the spin-triplet partial waves where OPE is attractive, pions do not
seem perturbative for
momenta $Q\simge 100$ MeV \cite{Fleming:1999ee}. Indeed, a careful
argument 
\cite{Birse:2005um} indicates that pions are non-perturbative
within the EFT regime in the
$^3S_1$-$^3D_1$, $^3P_0$, $^3P_2$-$^3F_2$, and possibly $^3D_2$
and $^3D_3$-$^3G_3$ waves. 
Thus, 
one should iterate OPE only in the waves where iteration is
needed, together with a chiral-invariant counterterm in each of
these waves \cite{Nogga:2005hy}. RGI, as well as a reasonable
description of data, can be maintained at higher orders as long as
these corrections are treated at the amplitude level in
distorted-wave perturbation theory
\cite{Valderrama:2009ei,Valderrama:2011mv,Long:2011qx,Long:2011xw,Long:2012ve},
as advocated in Refs.
\cite{Beane:2001bc,Nogga:2005hy,Birse:2007sx,Long:2007vp}.

Given the failure of NDA in the two-nucleon sector of Chiral EFT,
it is important to ascertain if few-body forces are not enhanced
as is the case for Pionless EFT 
\cite{Bedaque:1998kg,Bedaque:1998km,Bedaque:1999ve}. 
Under the assumption of NDA, three-body
forces appear only at subleading orders, and more-body forces
at even higher orders
\cite{Weinberg:1991um,Ordonez:1992xp,Weinberg:1992yk,vanKolck:1994yi}.
Virtually every application of ChPT in nuclear physics
\cite{Bedaque:2002mn,Epelbaum:2008ga,Machleidt:2011zz}
has assumed this to be appropriate.

Here we use RGI as a diagnostic tool for the appearance of
three-nucleon forces. 
A preliminary study of this
question was made in Ref. \cite{Nogga:2005hy} where the triton
binding energy was shown to approach a constant 
for cutoff values up to $4$ GeV
with a separable momentum regulator of the form 
$f_4(p/\Lambda)=\exp[-p^4/\Lambda^4]$, after
the two-nucleon problem was properly renormalized. 
We extend the LO cutoff dependence analysis of
Nogga {\it et al.} \cite{Nogga:2005hy} significantly. For the same
regulator function, we extend the two-nucleon renormalization to
cutoff values as high as $10$ GeV. We also examine the effects of
a change in regulator function to
$f_2(p/\Lambda)=\exp[-p^2/\Lambda^2]$ and
$f_6(p/\Lambda)=\exp[-p^6/\Lambda^6]$. Our results for the triton
binding energy are consistent with those of Nogga {\it et al.}
and we find similar cutoff dependence for the quartet and doublet
neutron-deuteron scattering lengths. An analysis of the residual
cutoff dependence of two- and three-nucleon observables
confirms that RGI does not require an LO three-body force, but does
indicate the appearance of an NLO correction in the $^1S_0$
channel, as pointed out by Long and Yang \cite{Long:2012ve}. Then,
we extend our analysis to include this NLO correction.
Again we find convergence of three-nucleon observables
with increasing cutoff, which
suggests that the three-body force is at least N$^2$LO.
However, there is some indirect evidence that this three-body
force might be numerically significant.

Of course we can only calculate a limited set of three-nucleon
observables \footnote{From here on, unless noted otherwise,
by ``three-nucleon observables'' we mean the triton binding energy and
the quartet and doublet neutron-deuteron scattering lengths.} 
and cannot exclude the possibility that other 
observables show relevant cutoff dependence.
However, being sensitive to many interactions, the
observables we study are likely to be a good diagnostic of
incomplete renormalization.
Very recently Kievsky {\it et al.} \cite{Kievsky:2016kzb} have argued
that for consistency with Pionless EFT
a three-nucleon force should be included
at LO in Chiral EFT as well.
Although a higher-order interaction can always be promoted in order to
improve agreement with data, we find no renormalization rationale
for promotion of the three-nucleon force in Chiral EFT, as
in Pionless EFT \cite{Bedaque:1998kg,Bedaque:1998km,Bedaque:1999ve}.

This paper is organized as follows. In Sec. \ref{sec:form}, the
RGI formulation of Chiral EFT at LO and NLO is presented, together with
the details of our calculation. Our numerical results are shown in
Sec. \ref{sec:results}, and Sec. \ref{sec:Summary} offers a
summary and conclusions.

\section{Theoretical Framework \label{sec:form}}

An EFT is based on the most general Lagrangian (or Hamiltonian)
built from the relevant low-energy degrees of freedom,
which is constrained only by some assumed symmetries.
Here we are interested in typical momenta $Q\sim m_\pi$,
so we consider nucleons and pions (the lightest
nucleon excitations contributing only beyond NLO)
under the QCD symmetries, including an
$SU(2)_L\times SU(2)_R$ chiral symmetry explicitly broken by the quark masses.

The basic assumption of EFT is that dynamics at a
low-energy scale does not depend on detailed assumptions about the
short-range dynamics,
which is encrypted in the interaction strengths,
or low-energy constants (LECs).
The LECs depend on the regularization parameter $\Lambda$
so that observables, obtained from scattering amplitudes,
are insensitive to the arbitrary regularization function.
Here we denote the breakdown scale of the current version
of Chiral EFT as $M_{hi}\simle M_{QCD}$, and
we will be concerned with the LECs that encode
short-range two- and three-nucleon interactions.

Because the EFT includes an infinite number of interactions, it is imperative
to organize them
according to their importance to amplitudes.
We estimate the importance of contributions according to the power
of $Q/M_{hi}$, and refer to the relation between this power and
the interactions in the Lagrangian as power counting.
Power counting offers a rationale to truncate the expansion of observables
with a relative error $(Q/M_{hi})^n$, where $n\ge 1$ is the power of
the next corrections.
We use some observables to fix the dependence of the LECs on $\Lambda$;
these observables remain exactly cutoff independent.
The residual cutoff
dependence of other observables will be considered
small as long as it is of the form
$(Q/\Lambda)^n$. In this case, variation for the cutoff from
$M_{hi}$ to a much larger value gives an estimate of the overall systematic
error of the truncation.

\subsection{Leading order}
In the original power counting scheme of
Weinberg~\cite{Weinberg:1990rz}, the LO potential includes OPE
between two nucleons. 
Denoting the spin (isospin) of nucleon $i$ by $\vs_i/2$ ($\vec{\tau}_i/2$) 
and the momentum transfer by $\vq=\vp'-\vp$, where $\vp$ ($\vp'$) is the 
initial (final) relative momentum,
the expression of OPE potential in momentum space is written as
\beq
V_{1\pi}^{(0)}(\vp',\vp)=-\frac{1}{(2\pi)^3}\frac{g_A^2}{4f_\pi^2}
\vec{\tau}_1\cdot\vec{\tau}_2
\frac{\vs_1\cdot\vq\,\vs_2\cdot\vq}{\vq^2+m_\pi^2}\,
\label{eq:ope}
\eeq
where $g_A=1.29$ is the pion-nucleon axial coupling,
$f_\pi=92.4$ MeV is the pion decay constant, and $m_\pi=138$ MeV is
the pion mass.
This potential has central and tensor components.

NDA suggests that at the same order one should supplement the
$^1S_0$ and $^3S_1$ partial waves with contact interactions.
In contrast, the analysis of Ref. \cite{Nogga:2005hy} shows that RGI
requires the presence of contact interactions at LO in 
{\it all}
waves where the singular pion tensor force is attractive and
treated non-perturbatively. 
Still, as the centrifugal barrier
increases, the low-energy effects of the tensor force diminish,
and for large angular momentum $l$ OPE 
does not need 
to be iterated within the regime of validity of the EFT. 
Only now is it being investigated how the $l$ suppression 
compares with the $Q/M_{hi}$ expansion, and so far results
are limited to spin-singlet channels \cite{PavonValderrama:2016lqn}.
Since neither the $l$
suppression nor the value of $M_{hi}$ are known precisely, one is
not able to make a sharp separation between waves where OPE is
non-perturbative or perturbative. 

In this work we consider an
LO contact potential
\bea
V_{ct}^{(0)}(\vp',\vp)&=&
\frac{1}{(2\pi)^3}\left[ \tilde{C}_{^1S_0} P_{^1S_0}+
                 \tilde{C}_{^3S_1} P_{^3S_1}
               +C_{^3P_0} \, p'p \, P_{^3P_0}  +
               C_{^3P_2} \, p' p \, P_{^3P_2}
              \right.
\nonumber\\
&&          \left.     +C_{^3D_2} \, p'^{2}p^2 \, P_{^3D_2}
               +C_{^3D_3} \, p'^{2}p^2 \, P_{^3D_3} \right],
\label{Vct}
\eea
where $C_X=C_X(\Lambda)$ and $P_X$ are, respectively,
a LEC in and the projector onto partial wave $X$.
Several comments are in order:

\begin{itemize}
\item OPE has a chiral-invariant delta-function singularity in
coordinate space that contributes in both $^1S_0$ and $^3S_1$
waves. NDA prescribes chiral-symmetric contact interactions in
these waves at LO, and we absorb the OPE delta function in the
corresponding LECs. The remaining part of OPE in $^1S_0$ (a Yukawa function)
is not singular in itself, but together with the chiral-symmetric
contact interaction produces an $m_\pi^2 \ln \Lambda$ divergence,
which requires at least one chiral-breaking counterterm
\cite{Kaplan:1996xu,Beane:2001bc}. Although this interaction has a
different form than the chiral-symmetric contact interaction, since
it involves also pion fields, nevertheless it does not produce new
effects up to NLO in the processes of interest here,
{\it as long as the pion mass is fixed}. 
In Eq. \eqref{Vct} we denote  by
$\tilde{C}_{^1S_0}$, $\tilde{C}_{^3S_1}$ the original
chiral-invariant LECs together with the OPE delta-function
coefficient and (in $^1S_0$) the chiral-breaking LEC.

\item
Numerical experimentation \cite{Nogga:2005hy} and a semi-analytical
argument \cite{Birse:2005um} suggest that OPE is non-perturbative
in all $P$ waves. OPE is not singular in $^1P_1$ and the tensor force
is repulsive in $^3P_1$.
Thus, of the $P$ waves, only $^3P_0$ and $^3P_2$ require counterterms
$C_{^3P_0}$ and $C_{^3P_2}$, respectively, at LO
--- although from NDA they would be expected only at N$^2$LO.
In the uncoupled $^3P_0$ wave OPE is attractive and almost as
strong as in the coupled $^3S_1$-$^3D_1$ channels:
as we are going to see shortly,
without
counterterms spurious $^3P_0$ bound states start appearing for cutoff
values which are only slightly larger than the ones
spawning
spurious states in the $^3S_1$-$^3D_1$ coupled channels \cite{Nogga:2005hy}.
The behavior in the coupled $^3P_2$-$^3F_2$ channels is
similar to $^3S_1$-$^3D_1$, {\it i.e.} each of them contains one attractive
and one repulsive eigenchannel.
Spurious bound states
appear in the $^3P_2$-$^3F_2$ channels for much larger cutoffs, meaning
a weaker singular potential \cite{Nogga:2005hy}.

\item
The situation is less obvious for the $D$ waves. As in other
singlet channels, in $^1D_2$ OPE is not singular. In the uncoupled $^3D_2$
wave without counterterm, the spurious states appear after $^3P_0$
but before $^3P_2$-$^3F_2$
\cite{Nogga:2005hy}. In
contrast, the coupled $^3D_3$-$^3G_3$ channels are even less favorable to
spurious states than $^3P_2$-$^3F_2$ \cite{Nogga:2005hy,Epelbaum:2006pt}.
This is consistent with the smallness of the $^3D_3$ phase shift
at low energies.
Still, the estimates from
Ref. \cite{Birse:2005um} clearly suggest the presence of
non-perturbative pions in 
$D$ waves. To be conservative, we
treat OPE non-perturbatively in all $D$ waves, which requires the
additional LO counterterms $C_{^3D_2}$, $C_{^3D_3}$,
even though they would be expected only at N$^4$LO on the basis
of NDA.

\item Pion exchange is likely to be 
perturbative for $l\ge 3$ \cite{Kaiser:1997mw,Kaiser:1998wa}, 
where phase shifts are small. 
But if OPE is iterated, none
of these waves (including $^3F_4$-$^3H_4$ and $^3G_4$ where OPE's
tensor force is attractive) accommodates spurious bound states  in
the range of cutoff values we investigate below (up to $10$ GeV).
This is consistent with the corresponding weakness of OPE
\cite{Birse:2005um}. In the following, we study low-energy
neutron-deuteron scattering including neutron-proton partial waves
non-perturbatively up to total angular momentum $j=4$, as in
Ref.~\cite{Epelbaum:2006pt}.
We will show that including such high angular-momentum components
in the low-energy three-nucleon system is unnecessary.
By including them, we are merely keeping some contributions
of higher order that do not introduce harmful cutoff
dependence in the cutoff range we consider.

\end{itemize}

The solution
of the dynamical equations involving the above potentials,
which are highly singular,
requires regularization.
We replace the potential by a regularized one,
\beq
V^{(0)}(\vp',\vp)=V_{1\pi}^{(0)}(\vp',\vp)+V_{ct}^{(0)}(\vp',\vp)
\to
V^{(0)}_\Lambda(\vp',\vp)=f_n(\vp'/\Lambda)\, V^{(0)}(\vp',\vp)\,
f_n(\vp/\Lambda)
\eeq
with the regulator function $f_n(x)$ in form of exponentials,
\beq
f_n(x)=\exp\left(-x^n\right),\quad n=2,4,6.
\label{regs}
\eeq
This form is
easily decomposed into partial waves,
ensuring
no additional mixing among waves.
We will use the regulator
with $n=4$,
unless otherwise mentioned. Regulator (in)dependence will be discussed later.

We study the two-body system (neutron-proton scattering and deuteron bound
state)
by solving the LS equation for the LO $T$ matrix,
\beq 
T^{(0)}(\vp',\vp)=V^{(0)}_\Lambda(\vp',\vp) +\int d^3 \vk \;
V^{(0)}_\Lambda(\vp',\vk) \frac{m_N}{\vp^2-\vk^2+i\epsilon}T^{(0)}(\vk,\vp), 
\label{LSeq}
\eeq
where $m_N=938.9$ MeV is the nucleon mass.
The on-shell amplitude $T(p)$ in a 
certain partial wave
then gives the LO phase shift in that wave,
\beq
\delta^{(0)}(p)=-\frac{i}{2}\ln\left[1-i\pi m_Np\, T^{(0)}(p)\right].
\label{LOphaseshift}
\eeq

There are many
ways to fix the values of counterterms at each cutoff. In this
work, we choose the following
ways to determine them:
\begin{itemize}
\item
All counterterms, except $C_{^3D_3}$, are fitted to PWA93
phase-shift data~\cite{Stoks:1993tb} at a 
laboratory energy 
$T_L=5$ MeV or 10 MeV.
In contrast, the ${}^3D_3$ phase
shifts are too small at low energies
to perform a reliable fit. Therefore for this wave we perform a global fit
of phase shifts up
to $T_L=200$
MeV through $\chi^2$ minimization, in a similar
way as it
was done in Ref.~\cite{Epelbaum:2006pt}.
\item
Alternatively, for $S$ waves ($^1S_0$ or $^3S_1$) we may adjust the
counterterms  to reproduce singlet ($a_{s}=-23.75$
fm) and triplet ($a_{t}= 5.42$ fm) scattering lengths.
Another 
option for the $^3S_1$ counterterm is to adjust it by fitting
the deuteron binding energy.
\end{itemize}
We will discuss the dependence on the fitting method in detail
in the next section. As we are going to see, renormalization
guarantees that no bound states cross the zero-energy threshold,
so phase shifts remain essentially cutoff independent. Instead,
deep bound states, which are beyond the region of validity of the
EFT, appear at certain cutoffs \cite{Nogga:2005hy}.

With the two-nucleon system properly renormalized,
the three-nucleon system is studied by
solving the Faddeev equations in configuration space with the LO
EFT potential. In particular, we calculate the triton binding
energy and neutron-deuteron scattering lengths.
The configuration-space two-nucleon potential is obtained by
carrying out a numerical
Fourier transformation of its momentum-space counterpart.
We remove the spurious deep
two-body bound states by using an orthogonalizing pseudo-potential
technique, see Refs. \cite{Kukulin:1978he,Nogga:2005hy}.
That is, we replace the two-body potential,
\beq
V^{(0)}_\Lambda \to {\tilde V}^{(0)}_\Lambda
= V^{(0)}_\Lambda + \sum_n |\psi_n\rangle
\lambda_n \langle \psi_n|,
\eeq
where the sum runs over the deep
states with wavefunctions $\psi_n$, and $\lambda_n$ are large
(positive) numbers. A three-body force is not included in these calculations;
instead, we
analyze the cutoff dependence of
three-body observables to check whether a three-body counterterm
is missing at LO.

The three-body wavefunction,
$\Psi^{(0)}$, is written in terms of Faddeev components, $\psi^{(0)}_{k}$,
\beq
\Psi^{(0)}(\vx,\vy)=\psi^{(0)}_1(\vx_1,\vy_1)+\psi^{(0)}_2(\vx_2,\vy_2)
+\psi^{(0)}_3(\vx_3,\vy_3),
\eeq
where a set of Jacobi coordinates, defined by
$\vx_{k}=(\vr_{j}-\vr_{i})\smallskip $
and
$\vy_{k}=(2\vr_{k}-\vr_{i}-\vr_{j})/\sqrt{3}$ with particle
indices $i,j,k=1,2,3$, is used.
Given isospin symmetry at LO,
the three Faddeev equations become formally identical,
and read
\begin{equation}
\left(E-H_{0}-{\tilde V}^{(0)}_{\Lambda ij}\right) \psi^{(0)}_{k}
={\tilde V}_{\Lambda ij}^{(0)}\left(\psi^{(0)}_{i}+\psi^{(0)}_{j}\right),
\label{EQ_FE}
\end{equation}
where $E$ is the three-body energy, $H_{0}$ is the three-particle
kinetic energy operator in the center-of-mass frame, 
and ${\tilde V}_{\Lambda ij}^{(0)}$ is the LO two-body interaction 
between particles $i$ and $j$. By using the
operator $P_{ij}$ for a permutation of particles $i$ and $j$,
the Faddeev components can also be written as
\begin{equation}
\psi^{(0)}_{i}+\psi^{(0)}_{j}=\left(P_{12}P_{23}+P_{23}P_{12}\right)\psi^{(0)}_{k}.
\end{equation}

The angular and spin-isospin dependence of the Faddeev components
is described using a bipolar harmonic basis, and the partial
Faddeev amplitude $F_\alpha(x_k,y_k)$ is defined from
\begin{equation}
\psi^{(0)}_{k}(\vx_{k},\vy_{k})=\sum\limits_{\alpha }
\frac{F_{\alpha }(x_{k},y_{k})}{x_{k}y_{k}}
\left\vert \left( l_{x}\left( s_{i}s_{j}\right) _{s_{x}}\right)
_{j_{x}}\left( l_{y}s_{k}\right) _{j_{y}}\right\rangle _{JM}\otimes
\left\vert \left( t_{i}t_{j}\right) _{t_{x}}t_{k}\right\rangle _{TT_{z}},
\label{EQ_FA_exp}
\end{equation}
where the index $\alpha$ represents all allowed combinations of the
quantum numbers present in the kets;
$s_{i}$ and $t_{i}$ are,
respectively, the spins and isospins of the individual particles;
$s_{x}$ and $t_{x}$ are,
respectively, the total spin and isospin of the two particles
associated with the Jacobi coordinate $x$;
$l_{x}$ and $l_{y}$ ($j_{x}$ and $j_{y}$) are the
orbital (total) angular momenta associated with the
corresponding Jacobi coordinates;
and $J$ and $M$ ($T$ and $T_z$) are, respectively,
the total angular momentum (isospin) and its third component.
For neutron-deuteron scattering at zero energy,
the Faddeev component
satisfies the following boundary condition
for $y_{nd}\rightarrow\infty$:
\begin{equation}
\label{eq:ndwave_norm}
\psi^{(0)}_{k_{nd}=0}(\vx_k,\vy_k)=
\sum\limits_{\alpha }\frac{\delta_{l_y,0}}{\sqrt{3}}
\left(1-2\frac{^{2J+1}a^{(0)}_{nd}}{\sqrt{3}y_k}\right) 
\frac{f^{d}_{\alpha}(x_{k})}{x_{k}}
\left\vert\left( l_{\alpha}\left( s_{i}s_{j}\right) _{s_{d}}\right)
_{j_{d}}\left( l_y s_{k}\right) _{j_{k}}\right\rangle _{JM}\otimes
\left\vert \left( t_{i}t_{j}\right)
_{t_{d}}t_{k}\right\rangle_{TT_{z}},
\end{equation}
where $f^{d}_{\alpha }$ is the deuteron's wavefunction component
for orbital angular-momentum component $l_{\alpha}$ dependent on
coordinate $\vx_k$; $s_{d}=1$, $j_{d}=1$, and $t_{d}=0$ are,
respectively, the deuteron's spin, angular momentum, and isospin;
and $^{2J+1}a^{(0)}_{nd}$ is the LO scattering length for either
doublet ($J=1/2$) or quartet ($J=3/2$) channels. Details of the
numerical methods employed here
can be found in
Ref.~\cite{lazauskas:tel-00004178,Lazauskas:2004hq,Song:2008zf}.

\subsection{Next-to-leading order}
By performing the renormalization procedure
the essential
cutoff dependence of observables is absorbed by counterterms. The effects
of the residual cutoff dependence
are comparable to the size of higher-order interaction terms.
According to the analysis performed by Long and Yang \cite{Long:2012ve},
the LO residual cutoff dependence in the $^1S_0$ partial wave
is $\propto Q/\Lambda$, and thus requires
a counterterm at ${\cal O}(Q/M_{hi})$.
This is consistent with our numerical analysis of
the residual cutoff dependence in $^1S_0$ phase shifts,
shown below.
The argument here is the same as used in Pionless EFT \cite{Bedaque:2002mn},
while NDA would assign this counterterm to N$^2$LO.

Therefore, for RGI, we
introduce a new counterterm in the $^1S_0$ channel at NLO,
which gives rise to a short-range contribution to the
effective range.
We write the NLO two-nucleon potential as
\beq
V_{ct}^{(1)}(\vp',\vp)=\frac{1}{(2\pi)^3}
\left[ {\tilde C}^{(1)}_{^1S_0}+{\tilde D}^{(1)}_{^1S_0}(p'^{2}+p^2) \right]
P_{^1S_0},
\label{NLOpot}
\eeq
where 
${\tilde D}^{(1)}_{^1S_0}$ is the new counterterm
needed at NLO, while
${\tilde C}^{(1)}_{^1S_0}$ is an NLO correction to the LO counterterm
$\tilde{C}^{(0)}_{^1S_0}$.
As before, we regulate the NLO potential,
\beq
V_{ct}^{(1)}(\vp',\vp)
\to
V^{(1)}_\Lambda(\vp',\vp)=f_n(\vp'/\Lambda)\, V_{ct}^{(1)}(\vp',\vp)\,
f_n(\vp/\Lambda),
\label{NLOpotreg}
\eeq
with the regulator functions \eqref{regs}.

While we compute the
LO $T$ matrix $T^{(0)}$ nonperturbatively by solving the LS equation with
the LO potential, we obtain the perturbative NLO correction $T^{(1)}$
using the distorted-wave Born approximation (DWBA)
\cite{Long:2012ve},
\bea
& &T^{(1)}(\vp',\vp)=V_\Lambda^{(1)}(\vp',\vp)
 \no
& &\quad
-\int d^3\vk \;
 \frac{m_N }{\vk^2-\vp^2-i\epsilon}
 \left[V_\Lambda^{(1)}(\vp',\vk)T^{(0)}(\vk,\vp)
 + T^{(0)}(\vp',\vk) V_\Lambda^{(1)}(\vk,\vp)
 \right]\no
& &\quad
+\int d^3\vk' \int d^3\vk \; T^{(0)}(\vp',\vk')
 \frac{m_N}{\vk'^{2}-\vp^2-i\epsilon} V_\Lambda^{(1)}(\vk',\vk)
 \frac{m_N}{\vk^{2}-\vp^2-i\epsilon} T^{(0)}(\vk,\vp).
\eea
The NLO correction to the $^1S_0$  phase shift,
$\delta^{(1)}$, is then obtained from the on-shell amplitude by
\beq
\delta^{(1)}(p)=-\frac{\pi m_Np}{2}
\mbox{Re}\left[e^{-2i\delta^{(0)}(p)}\, T^{(1)}(p)\right] ,
\label{NLOps}
\eeq
where
$\delta^{(0)}$ is the LO phase shift \eqref{LOphaseshift}.

For a given LO counterterm
$\tilde{C}^{(0)}_{^1S_0}$, the NLO counterterm
$\tilde{D}^{(1)}_{^1S_0}$ is determined by fitting the PWA93
phase shift at a different energy or, alternatively,
the effective range ($r_{s}= 2.77$ fm).
The NLO counterterm 
$\tilde{C}^{(1)}_{^1S_0}$ ensures that the NLO correction
vanishes for the observables used in fitting the LO counterterm.

NLO corrections to three-body observables are
calculated in
first-order perturbation theory. The correction to the
three-body binding energy is
\begin{equation}
\Delta E^{(1)}=\langle\Psi^{(0)}|\sum_{ij}V^{(1)}_{\Lambda ij}|\Psi^{(0)}\rangle,
\end{equation}
where $\Psi^{(0)}$ is the normalized three-body bound-state wavefunction due 
to LO potential.
Similarly, for the
correction to the scattering length,
\begin{equation}
\Delta a_{nd}^{(1)}
=\sqrt{\frac{3}{4}}\frac{m_N}{\hbar^2}
\langle\Psi^{(0)}_{k_{nd}=0}|\sum_{ij}V^{(1)}_{\Lambda ij}|\Psi^{(0)}_{k_{nd}=0}\rangle,
\end{equation}
where $\Psi^{(0)}_{k_{nd}=0}$ is the LO three-body wavefunction for 
neutron-deuteron scattering at zero
energy which is normalized to have unit flux as Eq. \eqref{eq:ndwave_norm}. 

We emphasize that NLO is strictly treated as a
perturbation, {\it i.e.} one insertion of the NLO potential
\eqref{NLOpotreg}.
This is done regardless of the size of the NLO potential 
relative to LO,
which is not an observable since both potentials are singular and 
cutoff-dependent.
A perturbative calculation of course does not reflect
the results of an exact solution of the LS and Faddeev equations
with the sum of LO and NLO potentials.
An exact solution includes all insertions of the NLO potential
but not the associated counterterms, so it
leads,
in general, to a non-renormalizable amplitude. 
For example, two insertions of the NLO potential
is an N$^2$LO effect, whose renormalization requires an N$^2$LO
interaction with four powers of momenta. 
The latter is missing if we iterate the NLO potential.
In Pionless EFT, it can
be shown analytically that the two-body amplitude is not
renormalizable for positive effective range if the NLO potential
is treated exactly \cite{Phillips:1997xu}, a manifestation of
Wigner's bound \cite{Wigner:1955zz}. Thus renormalization at the
two-nucleon level forces us here, as in Pionless EFT
\cite{vanKolck:1998bw,Chen:1999tn,Vanasse:2013sda}, into a
perturbative evaluation of NLO corrections in few-body systems.

As in any perturbation theory, the DWBA calculation of the NLO $T$ matrix will
explicitly break the unitarity of the $S$ matrix. The definition of the phase 
shift we use (Eq. \eqref{NLOps}) is not unique.
However, the error coming from the breaking of unitarity,
or alternatively from different definitions of the NLO phase shift, 
can be considered an N$^2$LO effect.
The amount of this breaking is 
one of the components of our total NLO error. 
While the difference between LO and NLO seems considerable at the level of the 
$^1S_0$ phase shifts, we are going to see
only relatively minor changes in the three-body observables we calculate.
Were this not the case, we would have been forced to consider an even
larger departure from NDA: we would have to elevate $^1S_0$ range effects
to LO, which at present can be done without violation of RGI only
through a dibaryon field \cite{Long:2013cya}. 

\section{Results \label{sec:results}}

\subsection{Two-nucleon system at LO}

In order to understand the effects of the counterterms at LO
we study neutron-proton scattering phase shifts and
the deuteron binding energy.
Phase shifts and their cutoff dependence are
obtained by solving the LS equation in each partial wave with
the LO Chiral EFT potential as an input.
The results of Ref.~\cite{Nogga:2005hy} are reproduced with cutoffs up to $4$
GeV, and
the cutoff range is extended
to 10 GeV. Our results are similar to those reported in
Refs.~\cite{Epelbaum:2006pt,Machleidt:2009bh}.

\begin{figure}[tbp]
\centering
\includegraphics[width=1\linewidth]{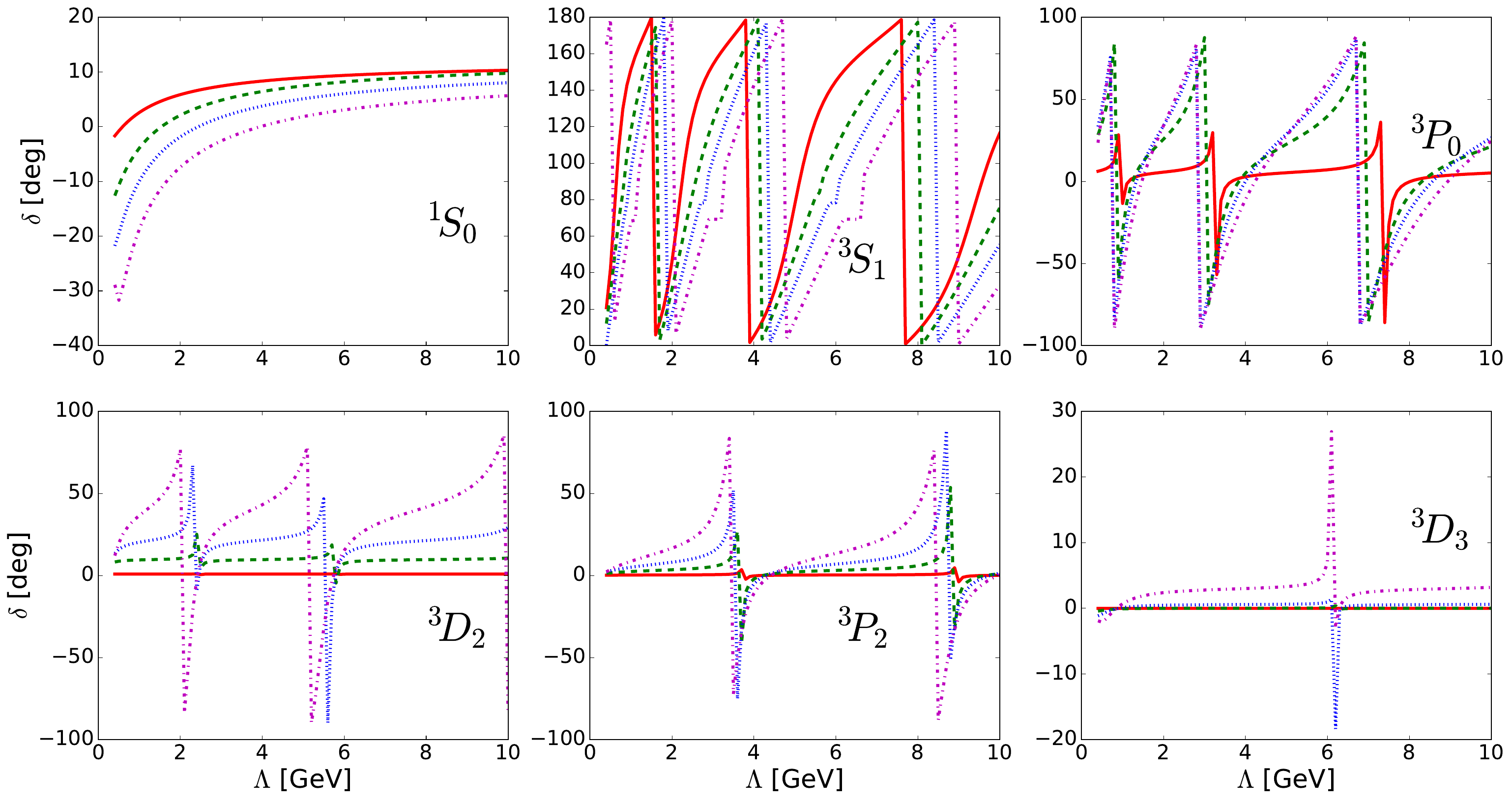}
\caption{(Color Online)
         Cutoff dependence of the phase shifts in $^1S_0$ and
         attractive tensor channels ($^3S_1$, $^3P_0$, $^3D_2$, $^3P_2$, $^3D_3$)
         without counterterms, with the  $n=4$ regulator.
         Results are given for different lab kinetic energies:
         10 MeV (red solid line), 50 MeV (green dashed line),
         100 MeV (blue dotted line), and 200 MeV (magenta dot-dashed line).}
\label{fig:fig_pwa_cut_nocounter}
\end{figure}

\begin{figure}[tbp]
\centering
\includegraphics[width=1\linewidth]{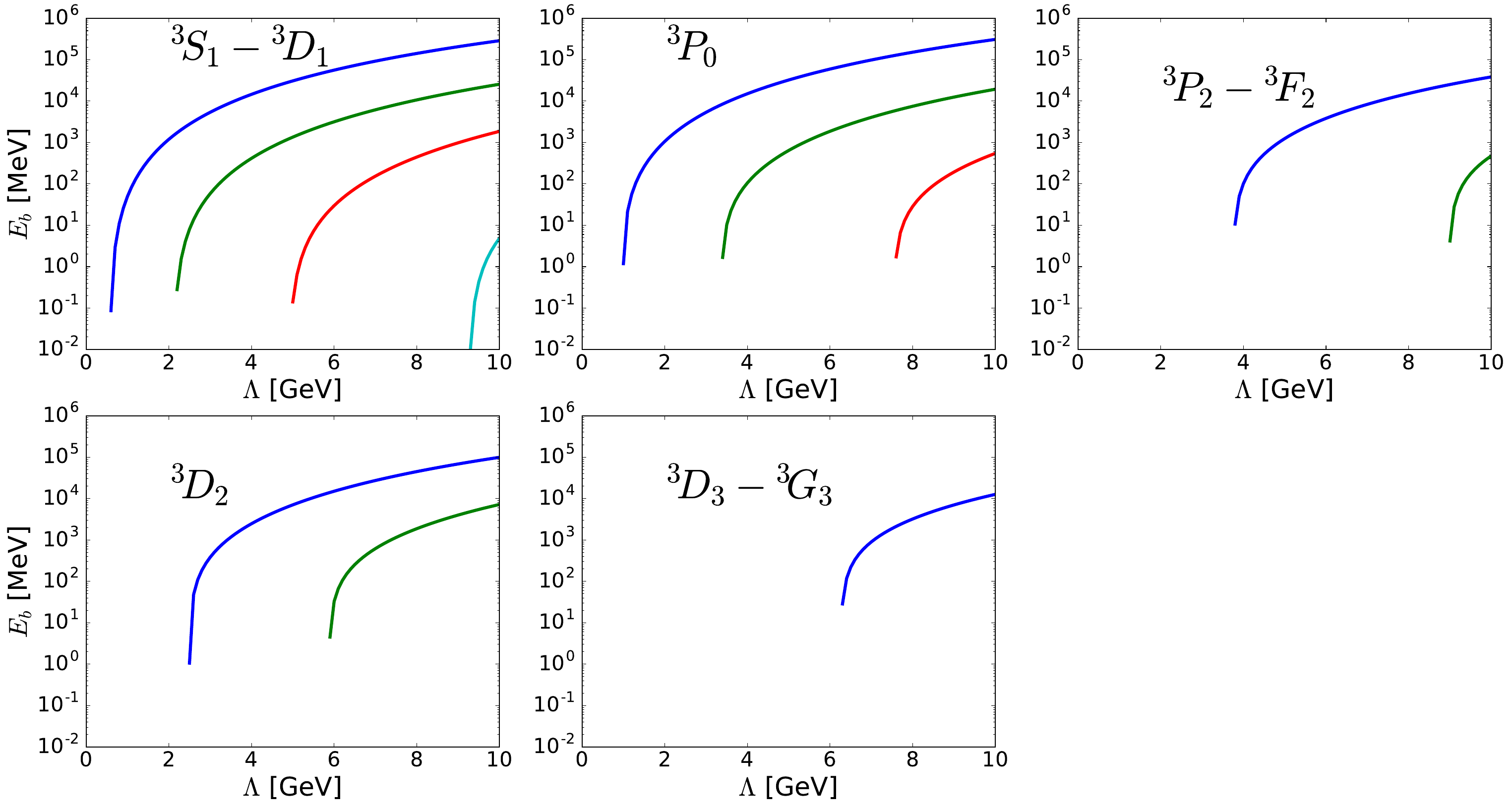}
\caption{Cutoff dependence of the binding energies of unphysical bound states
       in attractive tensor channels
       without counterterms, with the $n=4$ regulator.}
\label{fig:fig_deu_cut_nocounter}
\end{figure}

We plot the cutoff dependence of selected $np$ phase shifts
in Fig.~\ref{fig:fig_pwa_cut_nocounter} for a pure OPE potential,
with the $n=4$ regulator \eqref{regs} but
without any counterterm. The $^1S_0$ phase shift is
typical of other singlet channels in that it converges as
the cutoff increases. Phase shifts in
triplet channels where the OPE tensor force
is repulsive also converge, but in those where the tensor
force is attractive ($^3S_1$, $^3P_0$, $^3P_2$, $^3D_2$, $^3D_3$)
oscillations covering the range of phase shift values
are seen. These oscillations are a reflection of bound states
crossing threshold, as observed in Fig.~\ref{fig:fig_deu_cut_nocounter}
where the cutoff dependence of 
binding energies is displayed.

As noted in Ref. \cite{Nogga:2005hy},
the appearance of unphysical
bound states is due to the singular nature of the OPE potential in the
attractive tensor channels. To achieve
RGI one is obliged to
introduce counterterms at LO in these channels.
In $^3S_1$, the counterterm is the one prescribed by NDA,
but counterterms are needed in all attractive tensor channels
where pions are treated non-perturbatively.
Other channels do not contain spurious bound states and reveal
moderate cutoff dependence in phase shifts up to $\Lambda\sim 10$ GeV.
Nevertheless, in $^1S_0$ a counterterm is suggested by NDA and
should also be included, since there seems to be no argument
for its demotion from LO.

\begin{figure}[tbp]
\centering
\includegraphics[width=1\linewidth]{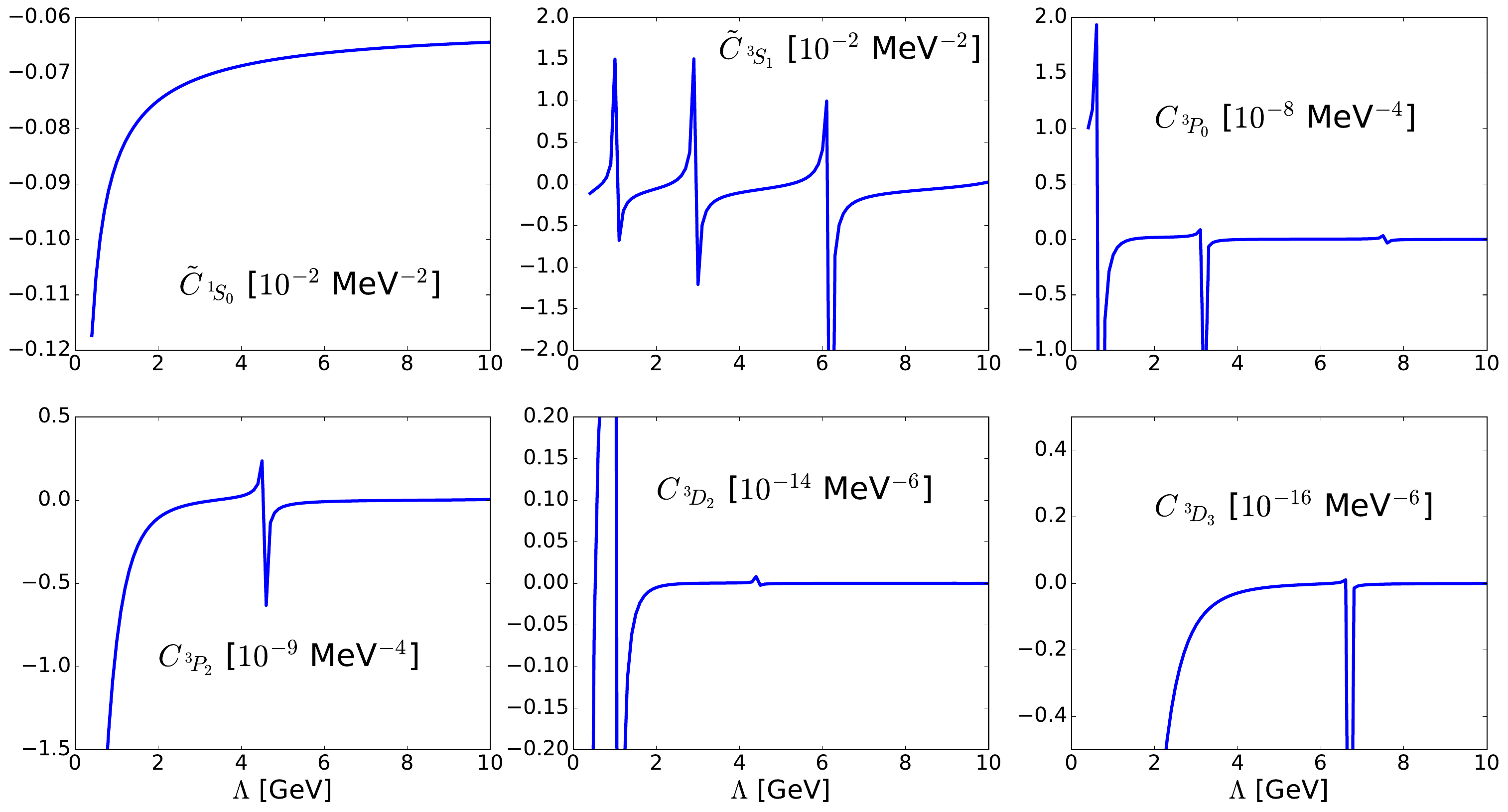}
\caption{Cutoff dependence of LO counterterms with the $n=4$ regulator.
        Counterterms are fitted to
        PWA93 phase shifts at 10 MeV, except for the $^3D_3$ channel which is
        globally fitted to phase shifts up to $200$ MeV.}
\label{fig:fig_pwa_ct}
\end{figure}

Once the counterterms in attractive tensor channels are included at LO,
they can be fitted to reproduce PWA93 phase shifts
at low energy.
In Fig.~\ref{fig:fig_pwa_ct} the cutoff dependence of the counterterms,
which are fitted to PWA93 at a laboratory energy of 10 MeV
(except for the $^3D_3$ channel which is fitted globally),
is presented.

\begin{figure}[tbp]
\centering
\includegraphics[width=1\linewidth]{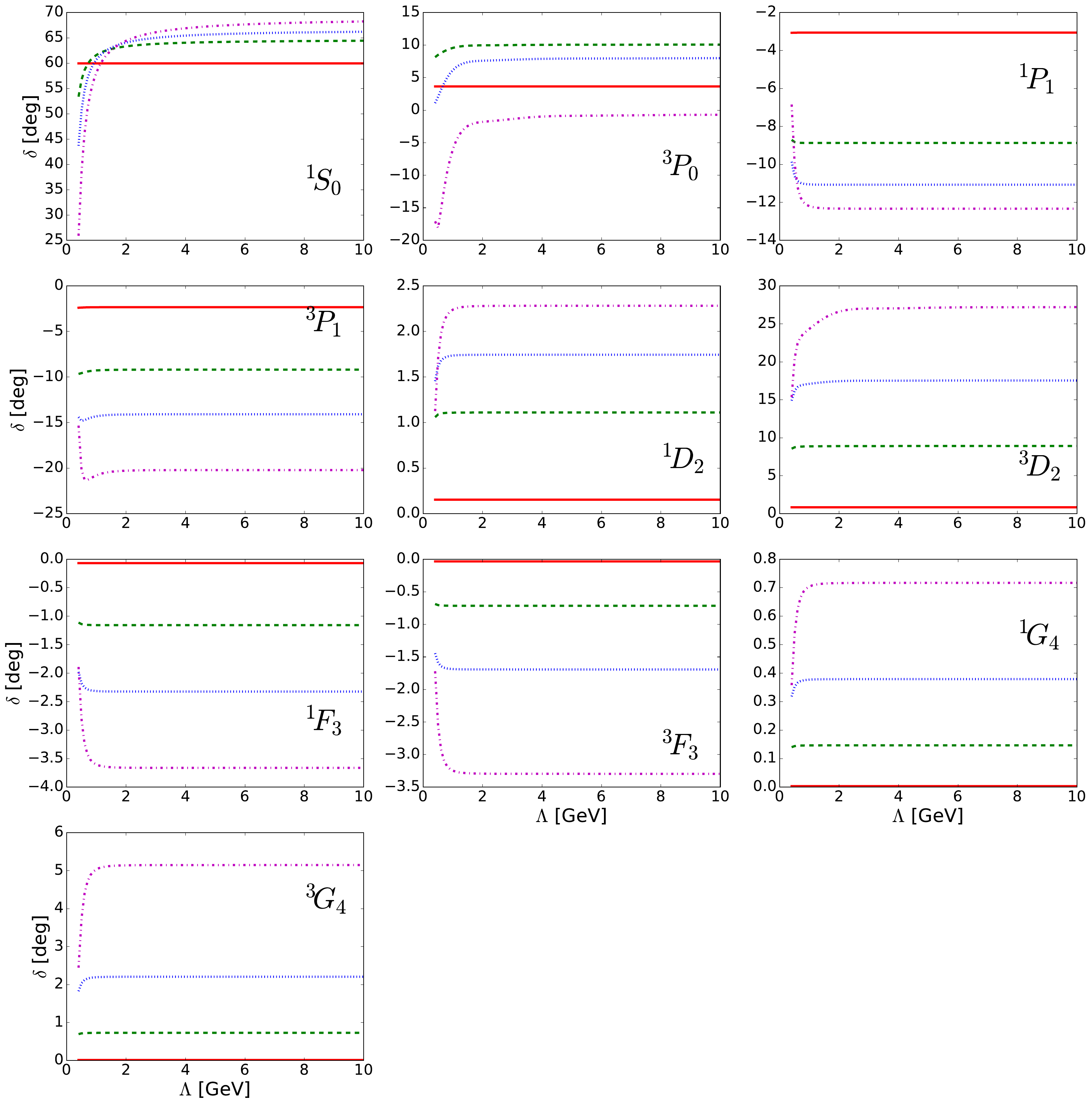}
\caption{(Color Online)
         Cutoff dependence of LO phase shifts in uncoupled channels up to total
         angular momentum $j=4$, with the $n=4$ regulator.
         Curves as in Fig. \ref{fig:fig_pwa_cut_nocounter}.
}
\label{fig:fig_pwa_cut_singlet_all}
\end{figure}

\begin{figure}[tbp]
\centering
\includegraphics[width=1\linewidth]{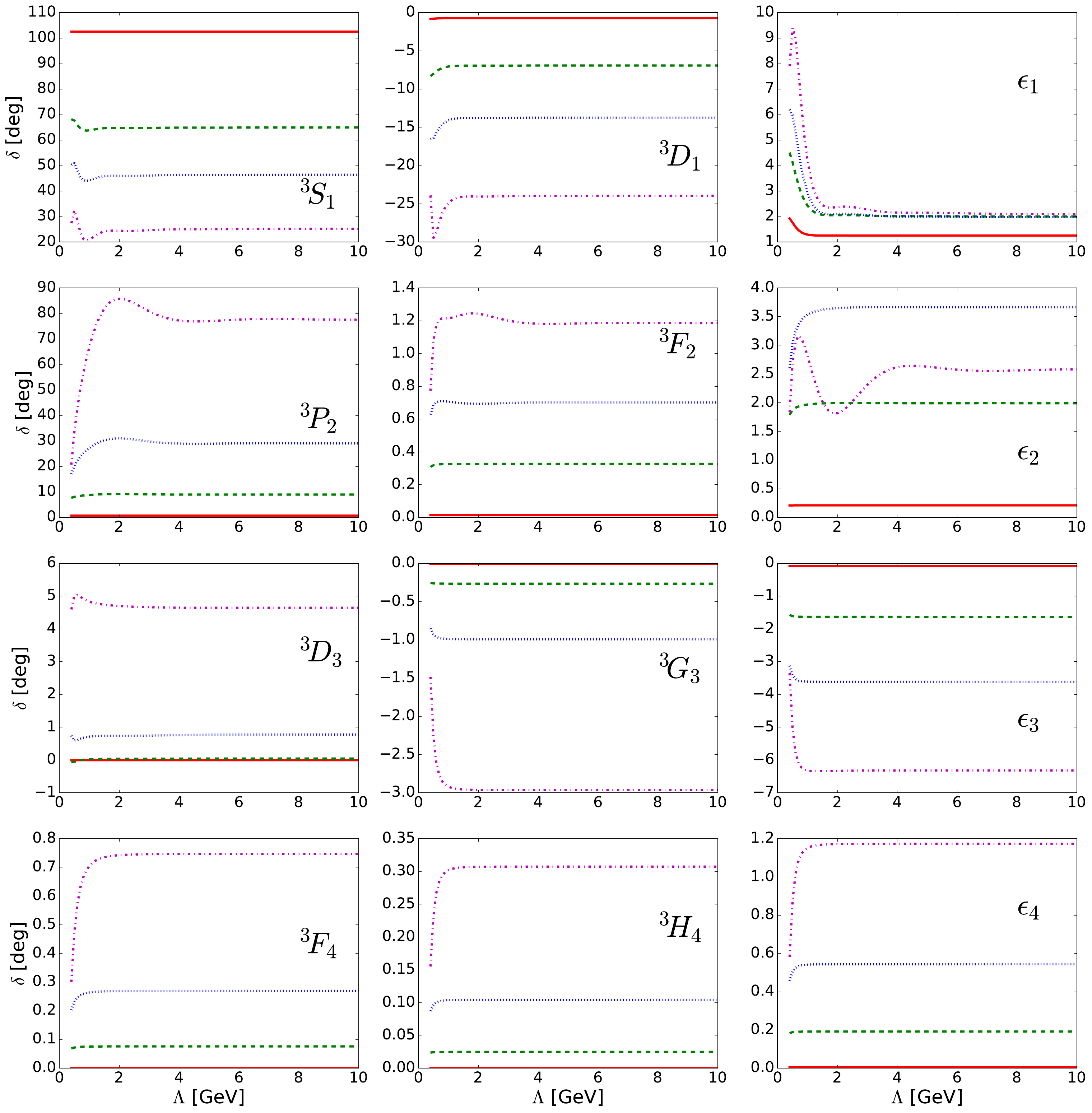}
\caption{(Color Online)
         Cutoff dependence of LO phase shifts in coupled channels up to total
         angular momentum $j=4$, with the $n=4$ regulator.
         Curves as
         in Fig. \ref{fig:fig_pwa_cut_nocounter}.
}
\label{fig:fig_pwa_cut_coupled_all}
\end{figure}

Because the
counterterms absorb most of the cutoff dependence in phase shifts,
we now observe convergence
at large cutoff values.
Figures \ref{fig:fig_pwa_cut_singlet_all}
and \ref{fig:fig_pwa_cut_coupled_all} show
the residual cutoff dependence of phase shifts in uncoupled
and coupled channels, respectively.
As expected in an EFT, the residual cutoff dependence
is largest at 
the largest energies.
For cutoff values larger than $2$ GeV
the bulk of the low-energy phase shifts becomes
cutoff independent. 
The mixing angle $\epsilon_2$
at $T_L=150$ MeV retains the strongest cutoff dependence,
nevertheless it becomes pretty small
for $\Lambda \simge 4$ GeV.
Note that the $^3F_4$-$^3H_4$ coupled channels,
dominated by the strong centrifugal barrier, do not require any
counterterm up to $\Lambda\sim 10$ GeV:
they show 
essentially no cutoff dependence, regardless of the fact that it is
an attractive tensor channel.
Of course, were the cutoff to be increased further, cutoff dependence would 
eventually appear. To investigate observables at such cutoff values one should
include another LEC or, more appropriately,
treat OPE in these channels as a subleading correction, that is, in DWBA.

\begin{figure}[tbp]
\centering
\includegraphics[width=1\linewidth]{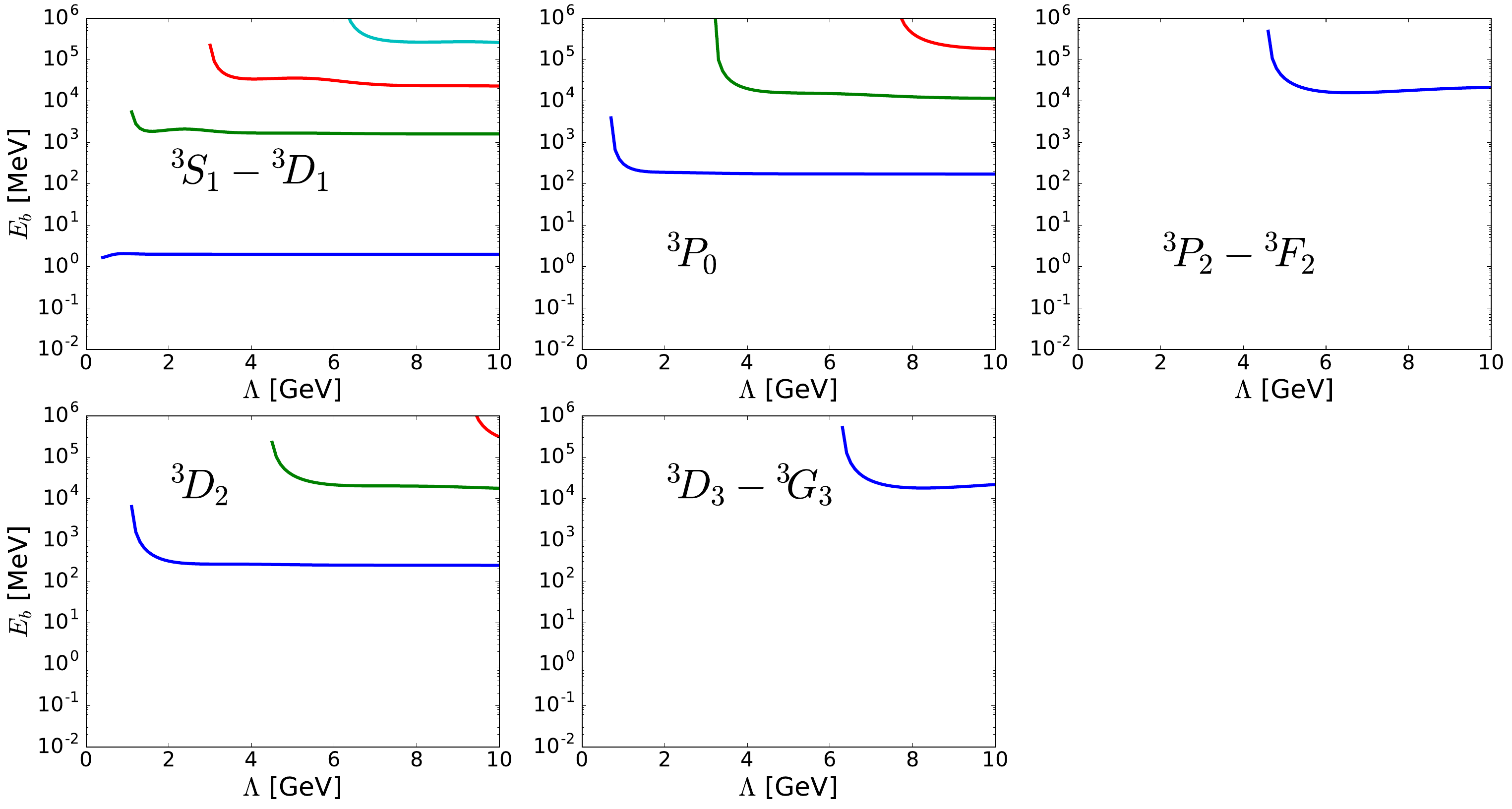}
\caption{Cutoff dependence of binding energies in attractive tensor channels
         after including a counterterm in each channel,
         with the $n=4$ regulator.
}
\label{fig:fig_deu_cut_counting2}
\end{figure}

After the renormalization procedure, the cutoff dependence of the
binding energies of spurious bound states also changes completely,
see Fig.~\ref{fig:fig_deu_cut_counting2}.
Only a single low-energy bound state appears, the deuteron
in the $^3S_1$ channel, and its binding energy is nearly cutoff
independent. Deep bound states exist, which also converge
as the cutoff increases, but they correspond
to states
outside the applicability of EFT.
These unphysical states must be removed when considering the
three-nucleon problem.
Turning the binding energy of the shallowest of these states,
the least-bound $^3P_0$ state ($\approx 170$ MeV), into an estimate of the
breakdown scale, we find $M_{hi}\sim 400$ MeV.
This is somewhat low, but a better
estimate requires a careful study of the convergence of observables
with order in the EFT expansion.

\begin{figure}[tbp]
\centering
\includegraphics[width=1\linewidth]{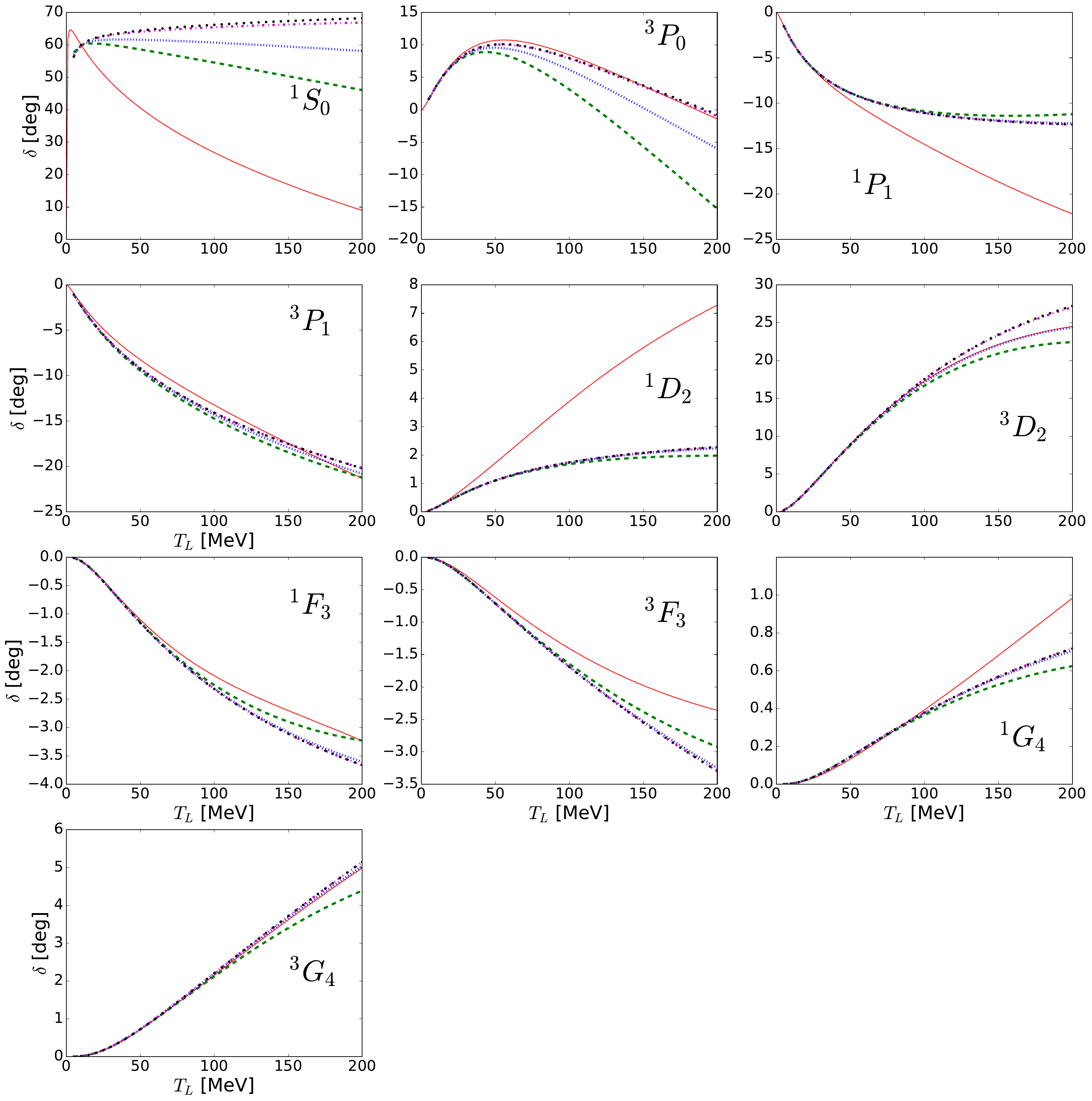}
\caption{(Color Online) Lab energy dependence of LO phase shifts
     in uncoupled channels
     up to $j=4$, with the $n=4$ regulator.
     Results are given for cutoff values of
     600 MeV (green dashed line), 1 GeV (blue dotted line),
     4 GeV (magenta dot-dashed line), and 10 GeV (black dots),
     and compared with PWA93 data (red solid line).
}
\label{fig:fig_pwa_energy_singlet_all}
\end{figure}

\begin{figure}[tbp]
\centering
\includegraphics[width=1\linewidth]{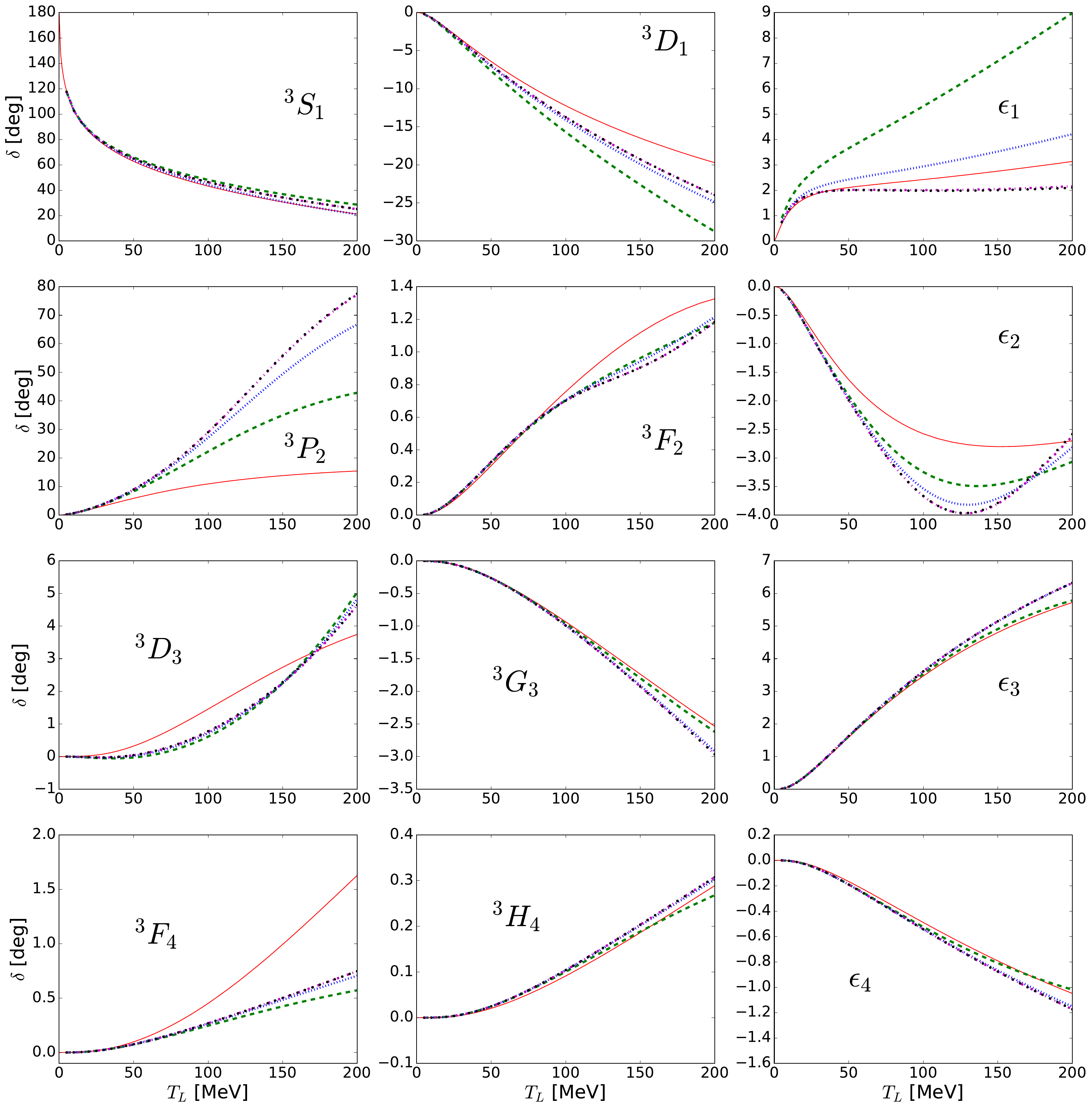}
\caption{(Color Online) Lab energy dependence of LO phase shifts
     in coupled channels
     up to $j=4$, with the $n=4$ regulator.
     Curves as in Fig. \ref{fig:fig_pwa_energy_singlet_all}.
     }
\label{fig:fig_pwa_energy_triplet_all_2}
\end{figure}

The energy dependence of the LO phase shifts in each partial wave is shown
in Fig.~\ref{fig:fig_pwa_energy_singlet_all} for uncoupled
channels and in Fig.~\ref{fig:fig_pwa_energy_triplet_all_2} for
coupled channels. The largest deviation in
comparison with PWA93 is
in the $^1S_0$ partial wave, for which
corrections appear at NLO.
As we
will see later, this deviation is indeed mitigated at NLO.

\begin{figure}[tbp]
\centering
\includegraphics[width=1\linewidth]{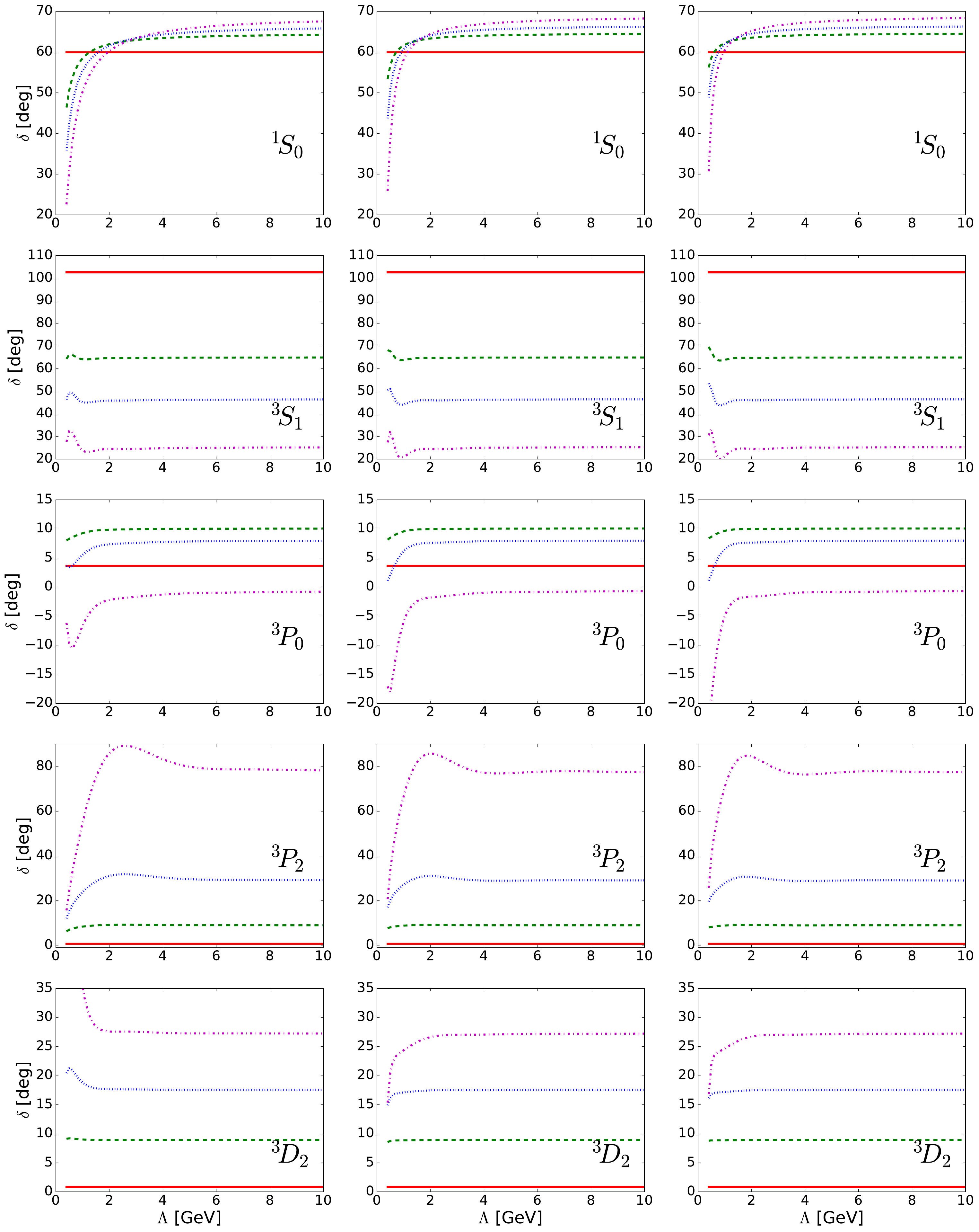}
\caption{(Color Online) Cutoff dependence of selected LO phase shifts for different regulators.
         Left, middle and right graphs correspond to regulators
         in Eq. \eqref{regs} with, respectively, $n=2,4,6$.
         Curves as in Fig. \ref{fig:fig_pwa_cut_nocounter}.
}
\label{fig:fig_2b_reg_all2}
\end{figure}

RGI requires not only independence of observables on the numerical
value of the cutoff $\Lambda$ but also independence on
the form of the regulator function itself.
In Fig.~\ref{fig:fig_2b_reg_all2} the cutoff dependence of phase
shifts is compared for the regulator functions $f_n(x)$ in Eq. \eqref{regs}
with $n=2,4,6$.
Dependence on the regulator function becomes negligible for large cutoffs 
but, as expected, it is still relevant at small cutoff values. 
The regulator-function dependence is in all cases not larger than
the $\Lambda$ variation for each regulator in the region 
$\Lambda \simge 1$ GeV.
Our results demonstrate that at large cutoff values
phase shifts become essentially independent of the choice of 
regulator function.

More generally, 
RGI ensures that physical observables are insensitive to the
arbitrary separation of long-range and short-range dynamics.
Because the long-range part of the potential is dominated by the OPE
interaction, we can expect insensitivity to details of the
fitting procedure, as long as they are used to reproduce low-energy
observables. In this work, we
check the sensitivity on the
fitting procedure by comparing results with counterterms
that are fitted at different energies, with the $n=4$ regulator.
Table~\ref{tbl:counter_term_fit} lists our four fitting choices,
labeled (I), (II), (III), and (IV).

\begin{table}
\begin{tabular}{c||c|c|c||c}
index & $^1S_0$ LO &  $^3S_1$ LO &  
$l>0$ LO &  $^1S_0$ NLO  \\
\hline
(I)     &$\delta_{10}$&$\delta_{10}$&$\delta_{10}$&$\delta_{10}$,$\delta_{20}$ \\
(II)    &$\delta_{5}$ &$\delta_{5}$ &$\delta_{5}$ &$\delta_{5}$,$\delta_{10}$ \\
(III)   &$a_{s}$  &$a_{t}$  &$\delta_{10}$& $a_{s}$, $r_{s}$  \\
(IV)    &$a_{s}$  &$E_d$        &$\delta_{5}$ & $a_{s}$, $r_{s}$ \\
\end{tabular}
\caption{The four choices of fitting
procedure employed in this work
at LO and NLO,
for $^1S_0$, $^3S_1$, and 
$l>0$ counterterms.
$\delta_{E}$ represents the PWA93 phase shift at kinetic energy $E$ MeV;
$a_{s,t}$ ($r_s$) stands for the singlet/triplet
scattering lengths (singlet effective range);
and $E_d$ is deuteron binding energy.
In all cases the $^3D_3$ counterterm is determined by a global
$\chi^2$ minimization for phase shifts up to 200 MeV.
}
\label{tbl:counter_term_fit}
\end{table}

\begin{figure}[tbp]
\centering
\includegraphics[width=0.8\linewidth]{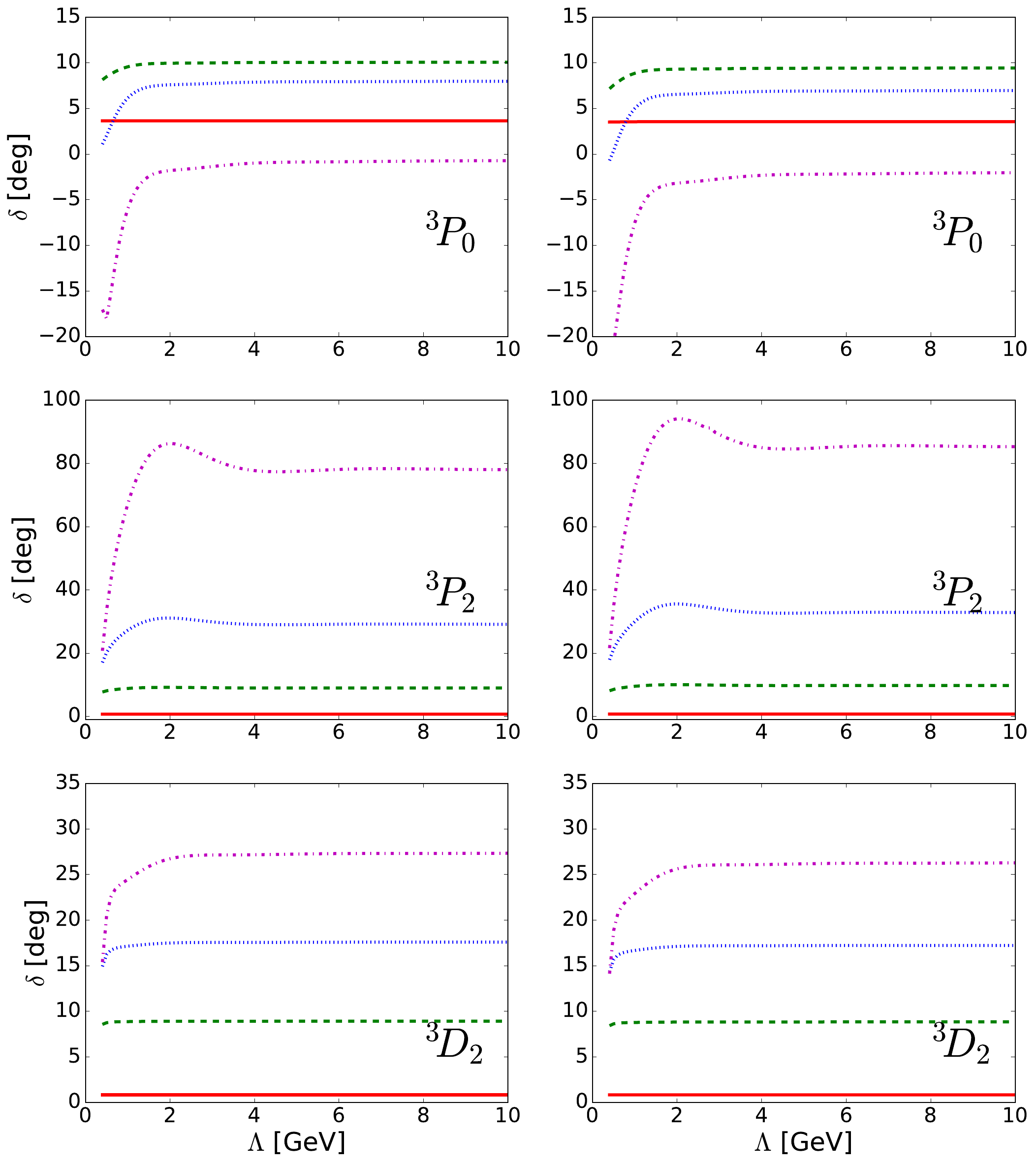}
\caption{(Color Online) Cutoff dependence of LO phase shifts
         for 
         $l>0$ partial waves with the $n=4$ regulator,
         for different fitting energies.
         The graphs on the left column are obtained with counterterms
         fitted at $T_L=10$ MeV and those on the right column,
         with counterterms fitted at $T_L=5$ MeV.
         Curves as in Fig. \ref{fig:fig_pwa_cut_nocounter}.
}
\label{fig:fig_2b_pwa_fitenergy_pd}
\end{figure}

\begin{figure}[tbp]
\centering
\includegraphics[width=0.8\linewidth]{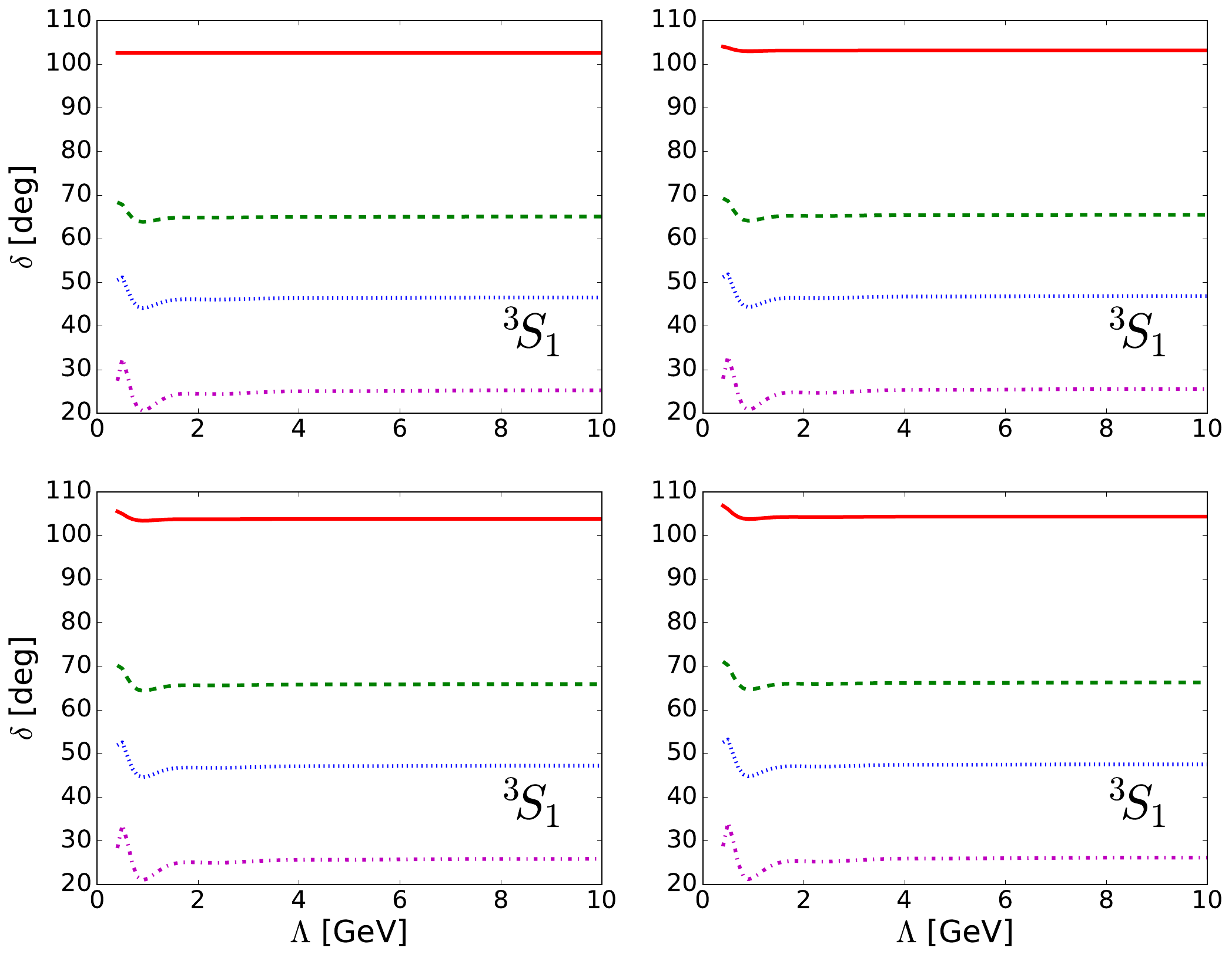}
\caption{(Color Online) Cutoff dependence of the LO $^3S_1$ phase shift
         with the $n=4$ regulator, for different fitting methods.
         Graphs are obtained with counterterms fitted to PWA93 phase shifts at
         10 MeV (upper left) or 5 MeV (upper right),
         to the scattering length (lower left), and
         to the deuteron binding energy (lower right).
         Curves as in Fig. \ref{fig:fig_pwa_cut_nocounter}.
}
\label{fig:fig_2b_pwa_fitenergy_3s1}
\end{figure}

\begin{figure}[tbp]
\centering
\includegraphics[width=1\linewidth]{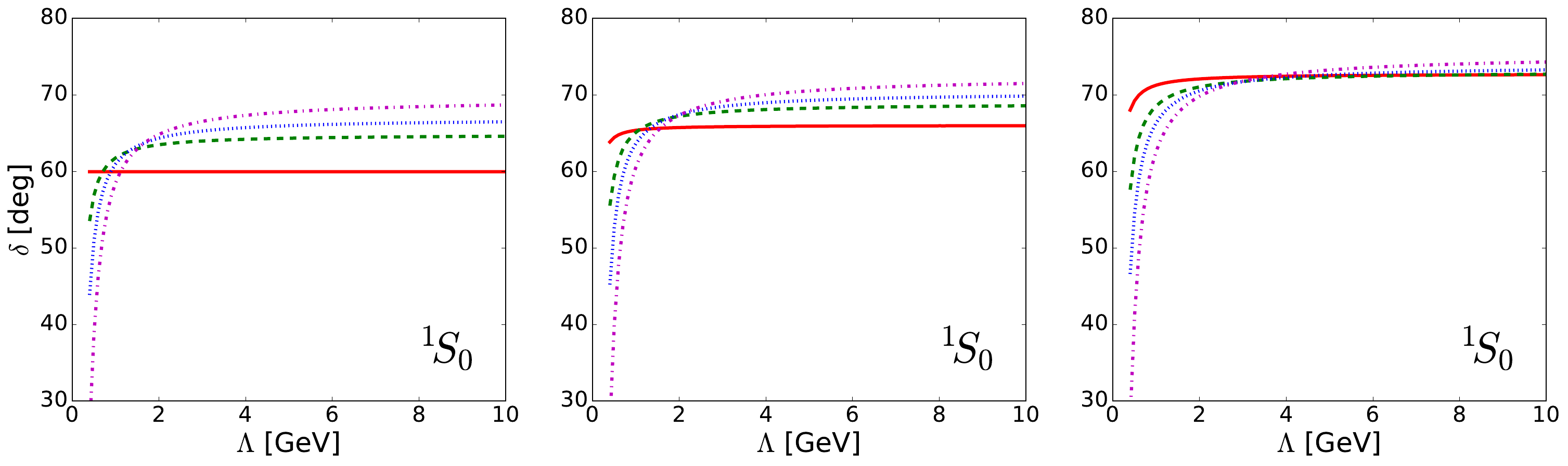}
\caption{(Color Online) Cutoff dependence of the LO $^1S_0$ phase shift
         with the $n=4$ regulator, for different fitting methods.
         Graphs are obtained with counterterms fitted to
         PWA93 phase shifts at 10 MeV (left) or 5 MeV (middle),
         and to the scattering length (right).
         Curves as in Fig. \ref{fig:fig_pwa_cut_nocounter}.
}
\label{fig:fig_2b_pwa_fitenergy_1s0}
\end{figure}

The $^3P_0$, $^3P_2$ and $^3D_2$ counterterms are fitted to
PWA93 phase shifts at $T_L=5$ MeV or 10 MeV.
Figure~\ref{fig:fig_2b_pwa_fitenergy_pd}
shows a comparison of the cutoff dependence for 
$l>0$ phase shifts
at several energies,
for fitting procedures (I) and (II).
The $^1S_0$ and $^3S_1$ counterterms are fitted to phase shifts at
$T_L=5$ MeV or 10 MeV, or to scattering lengths
or (for $^3S_1$) the deuteron binding energy.
Figure~\ref{fig:fig_2b_pwa_fitenergy_3s1} shows the
$^3S_1$ phase shifts as a function of the cutoff
with counterterms fitted following strategies (I), (II), (III), and (IV),
while Fig.~\ref{fig:fig_2b_pwa_fitenergy_1s0} shows the
analogous results for the $^1S_0$ phase shifts
for strategies
(I), (II), and (III).
Though all cases show cutoff independence at large cutoff values,
the low-energy $^1S_0$ phase shifts are somewhat sensitive to
the fitting method.
This is because of the large deviation
of the LO phase shift from PWA93 data in $^1S_0$ even at low energies.
This uncertainty is expected to be reduced
once higher-order corrections are included.
For all other channels the fitting procedure has very little impact
on the phase shifts. This impact increases with energy but is not larger
than the cutoff variation for $\Lambda \simge 1$ GeV.

\begin{figure}[tbp]
\centering
\includegraphics[width=0.8\linewidth]{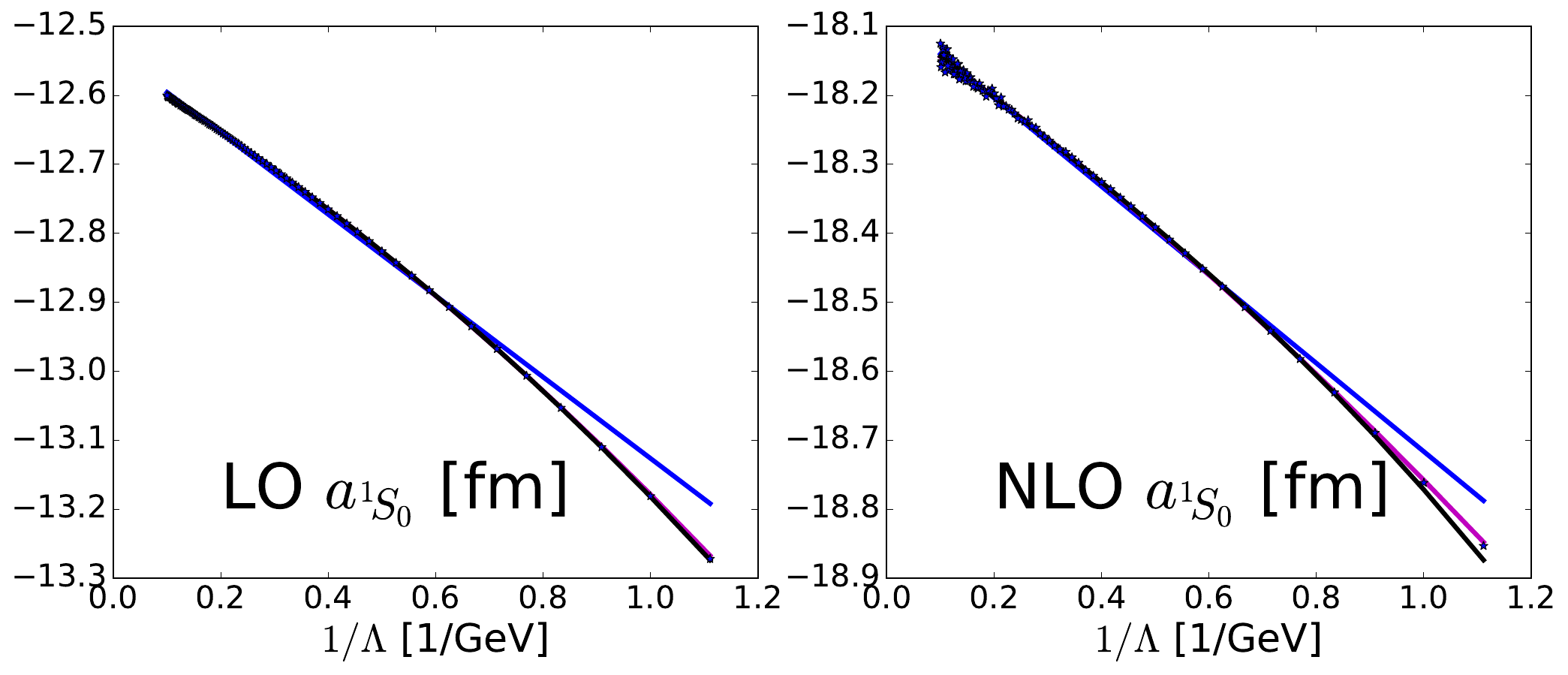}
\caption{(Color Online) Cutoff dependence of the $^1S_0$ scattering length
      at LO (left) and NLO (right) 
      with the $n=4$ regulator and fitting procedure (II).
      The blue, magenta, and black solid lines are, respectively, 
      two-, three-, and four-parameter fits with 
      Eqs. \eqref{1S0scattlengthfit:pol} and \eqref{1S0scattlengthfit:pol1} 
      to results for 
      $\Lambda \ge 1.2$ GeV. 
}
\label{fig:fit_2b_sc1s0:v2}
\end{figure}

The residual cutoff dependence may be related to the order of missing
higher-order corrections.
As an example, let us consider the cutoff dependence of the $^1S_0$ scattering
length
in the cutoff range $\Lambda\ge 1.2$ GeV for the $n=4$ regulator,
when the counterterm is fitted to the phase shift at $T_L=5$ MeV.
We fit the cutoff dependence with a power series
\bea 
a_{s}^{(0)}(\Lambda)= a_{s}^{(0)}(\infty)
 \left[1+\frac{p_{s1}^{(0)}}{\Lambda}
 +\left(\frac{p_{s2}^{(0)}}{\Lambda}\right)^2
 +\left(\frac{p_{s3}^{(0)}}{\Lambda}\right)^3   
 +\cdots
 \right],
\label{1S0scattlengthfit:pol}
\eea 
truncating the series at successively larger powers.
The fitting parameters $a_{s}^{(0)}(\infty)$ and $p_{s1,2,3}^{(0)}$
are summarized in Table \ref{tbl:1s0scattering:fit}
and the corresponding fits are shown in 
the left panel of Fig. \ref{fig:fit_2b_sc1s0:v2}.
(Results are similar for other cutoff ranges, regulators,
and fitting strategies.)
The parameters are relatively stable as the number of fit parameters
increases. 
For the four-parameter fit, the parameters
$a_{s}^{(0)}(\infty)$ and $p_{s1,2}^{(0)}$ have nearly converged (within errors),
while the large  $p_{s3}^{(0)}$ error suggests that fits with more parameters
are meaningless.

As expected, the momenta $p_{s1,2,3}^{(0)}$ are given by low-energy scales:
while $p_{s2,3}^{(0)}\sim m_\pi$,
$p_{s1}^{(0)}$ is somewhat smaller, possibly as a consequence 
of the fine tuning in this channel, but definitely non-vanishing.
The residual cutoff dependence $\propto \Lambda^{-1}$ is consistent with the 
argument in Ref. \cite{Long:2012ve},
which implies a $Q/M_{hi}$
counterterm. 
Also, note that with this input
the LO potential does not reproduce the scattering length
--- that is, the long-range component of the wavefunction --- well.
The introduction of an NLO
correction will give more accurate values 
for the scattering length and the effective range.

\begin{table}
\begin{tabular}{c|c|c|c|c|c|c|c}
     \multicolumn{4}{c|}{LO}  & \multicolumn{4}{c}{NLO}
\\
\hline    
     $a_s^{(0)}(\infty) $ &  $p_{s1}^{(0)}$  & $p_{s2}^{(0)}$ & $p_{s3}^{(0)}$ 
     & $a_s^{(1)}(\infty)$ &  $p_{s1}^{(1)}$  & $p_{s2}^{(1)}$ & $p_{s3}^{(1)}$
\\       
\hline    
     $-12.5$ & $47.0(0.3)$ & -  & - &
     $-18.1$ & $35.4(0.3)$ & -  & - 
\\
     $-12.6$ & $37.5(0.1)$ & $111(11)$  & - &
     $-18.1$ & $30.2(0.9)$ & $82.8(34)$  & - 
\\
     $-12.6$ & $38.3(0.3)$ & $100(28)$  & $124(86)$ &
     $-18.1$ & $32.5(2.8)$ & $8.1(88)$  & $176(183)$ 
\\    
\end{tabular}
\caption{Parameters of fits using Eqs. \eqref{1S0scattlengthfit:pol}
         and \eqref{1S0scattlengthfit:pol1} 
         to the leading-order and next-to-leading-order results 
         for the $^1S_0$ scattering length
         from fitting procedure (II) for cutoff values $\Lambda\geq 1.2$ GeV.
        The parameters $a_s^{(m)}(\infty)$ and $p_{s1,2,3}^{(m)}$ ($m=0,1$)
        are in fm and MeV, respectively.
        Numbers in parentheses represent fitting errors.  
}
\label{tbl:1s0scattering:fit}
\end{table}

In summary, our two-nucleon results are consistent with the results of
Ref.~\cite{Nogga:2005hy}, and also demonstrate that
regulator-independence 
is obtained up to cutoff values
as large as $10$ GeV for all partial waves.
We turn now to the LO cutoff dependence of
three-nucleon low-energy observables.

\subsection{Three-nucleon system at LO}

\begin{figure}[tbp]
\centering
\includegraphics[width=0.5\linewidth]{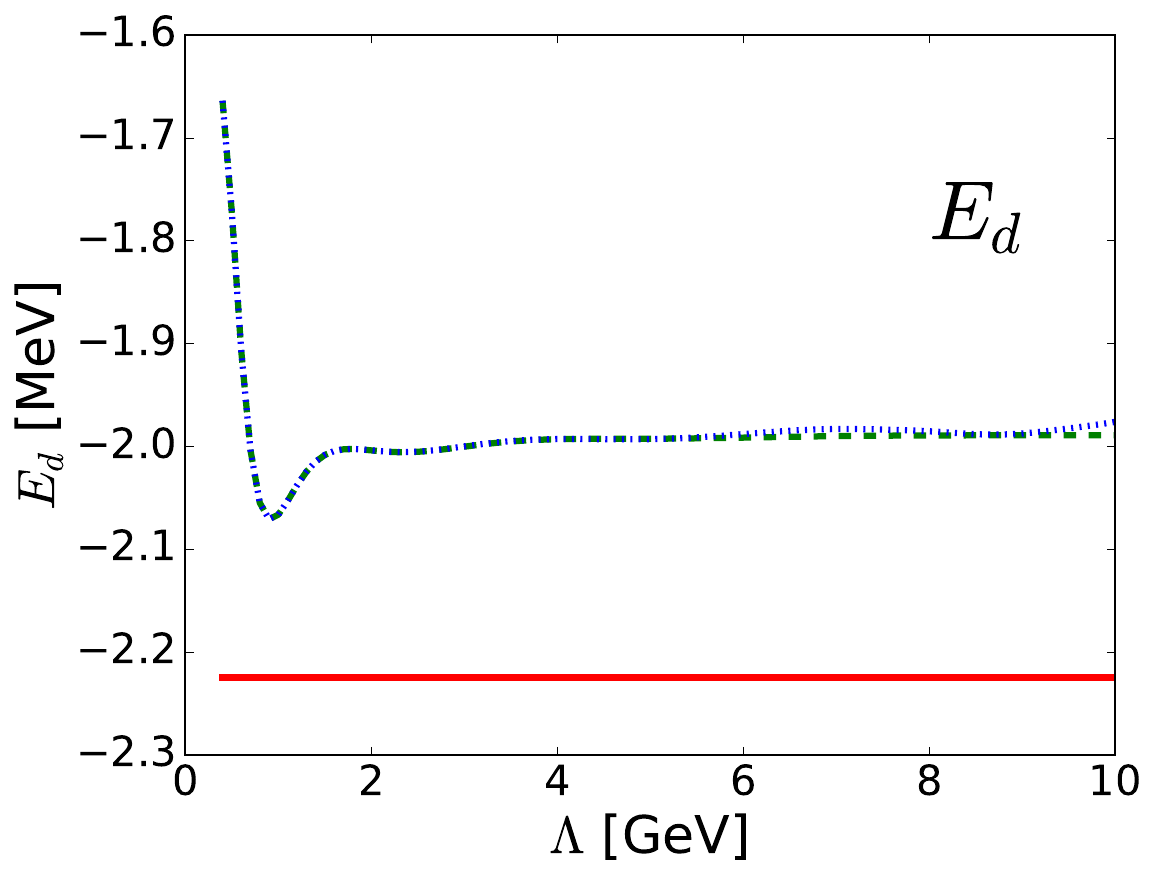}
\caption{(Color Online) Cutoff dependence of the LO deuteron energy
         with the $n=4$ regulator.
         Results are shown for momentum-space (green dashed line)
         and configuration-space (blue dotted line) calculations,
         in comparison with experiment (red solid line).
}
\label{fig:fig_3b_deu}
\end{figure}

In order to check the cutoff dependence of the three-nucleon system with
the LO Chiral EFT potential,
we calculate neutron-deuteron scattering lengths and
the triton binding energy by solving the Faddeev equation in
configuration space.
The two-nucleon EFT potential at
LO is Fourier-transformed numerically from momentum space.
There appears a small oscillating cutoff dependence
in three-body results at large cutoff values.
Similar oscillations also
show up in the deuteron binding energy when calculated in
configuration space, see Fig.~\ref{fig:fig_3b_deu}.
These oscillations are
absent if the deuteron binding energy is calculated directly in
momentum space,
and thus can be attributed to errors in
the numerical Fourier transformation.

\begin{figure}[tbp]
\centering
\includegraphics[width=1\linewidth]{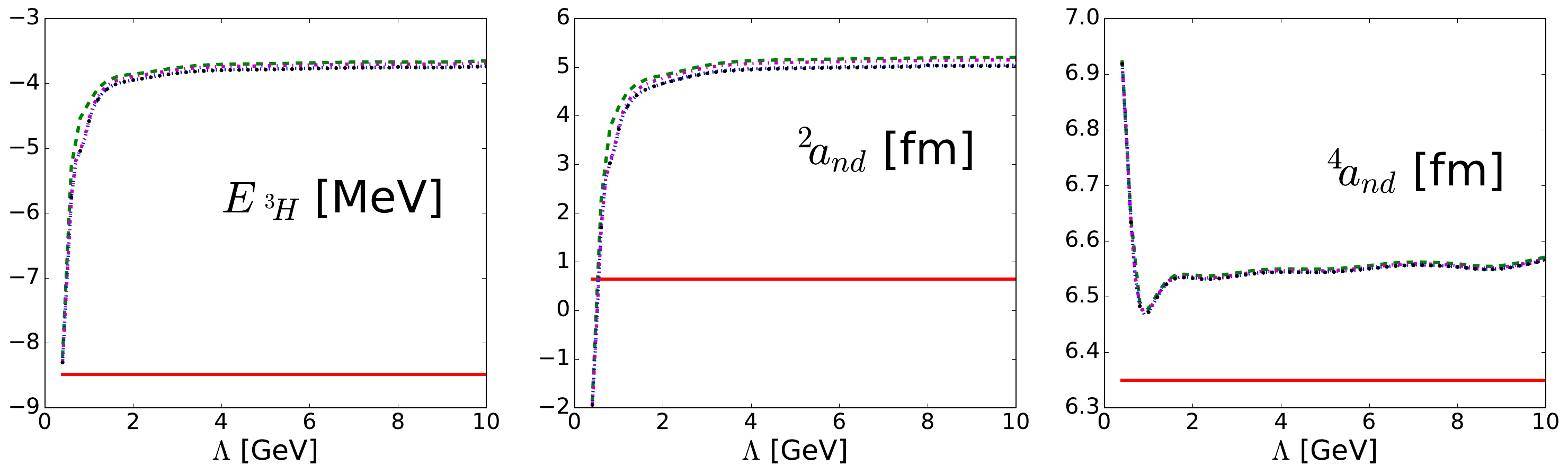}
\caption{(Color Online) Cutoff dependence of the triton energy (left), and
         $J=1/2$ (middle) and $J=3/2$ (right) neutron-deuteron
         scattering lengths at LO,
         for the $n=4$ regulator.
         Results are shown for
         various $j_{max}$ values in the two-nucleon interaction:
         $j_{max}=1$ (green dashed line),
         $j_{max}=2$ (blue dotted line),
         $j_{max}=3$ (magenta dot-dashed line),
         and $j_{max}=4$ (black dots), in comparison with
         experiment (red solid line).}
\label{fig:fig_3b_jmax}
\end{figure}

Figure~\ref{fig:fig_3b_jmax} shows the cutoff dependence of
the triton binding energy
and neutron-deuteron scattering lengths in doublet and
quartet channels, when they are computed using the LO potential
with counterterms fitted at $T_L=10$ MeV and the $n=4$ regulator.
To see the three-nucleon effects of two-nucleon
partial waves,
we introduce a maximum total two-nucleon angular momentum
$j_{max}$ in the two-nucleon interaction,
such that ${\tilde V}_{\Lambda ij}^{(0)}(\vx)=0$ if $j_x > j_{max}$.
Variation of $j_{max}$ from 1 to 4
indicates that
adding two-nucleon partial waves 
leads to small effects,
which are negligible for $j_{max}\geq 2$.
This is consistent with the perturbativeness of $l\ge 3$ two-nucleon
partial waves. For definiteness, below we show results for $j_{max}=4$.
We see that cutoff independence
is achieved in low-energy three-body observables:
they vary by less than about 10\%
for $\Lambda \simge 2$ GeV.

\begin{figure}[tbp]
\centering
\includegraphics[width=0.8\linewidth]{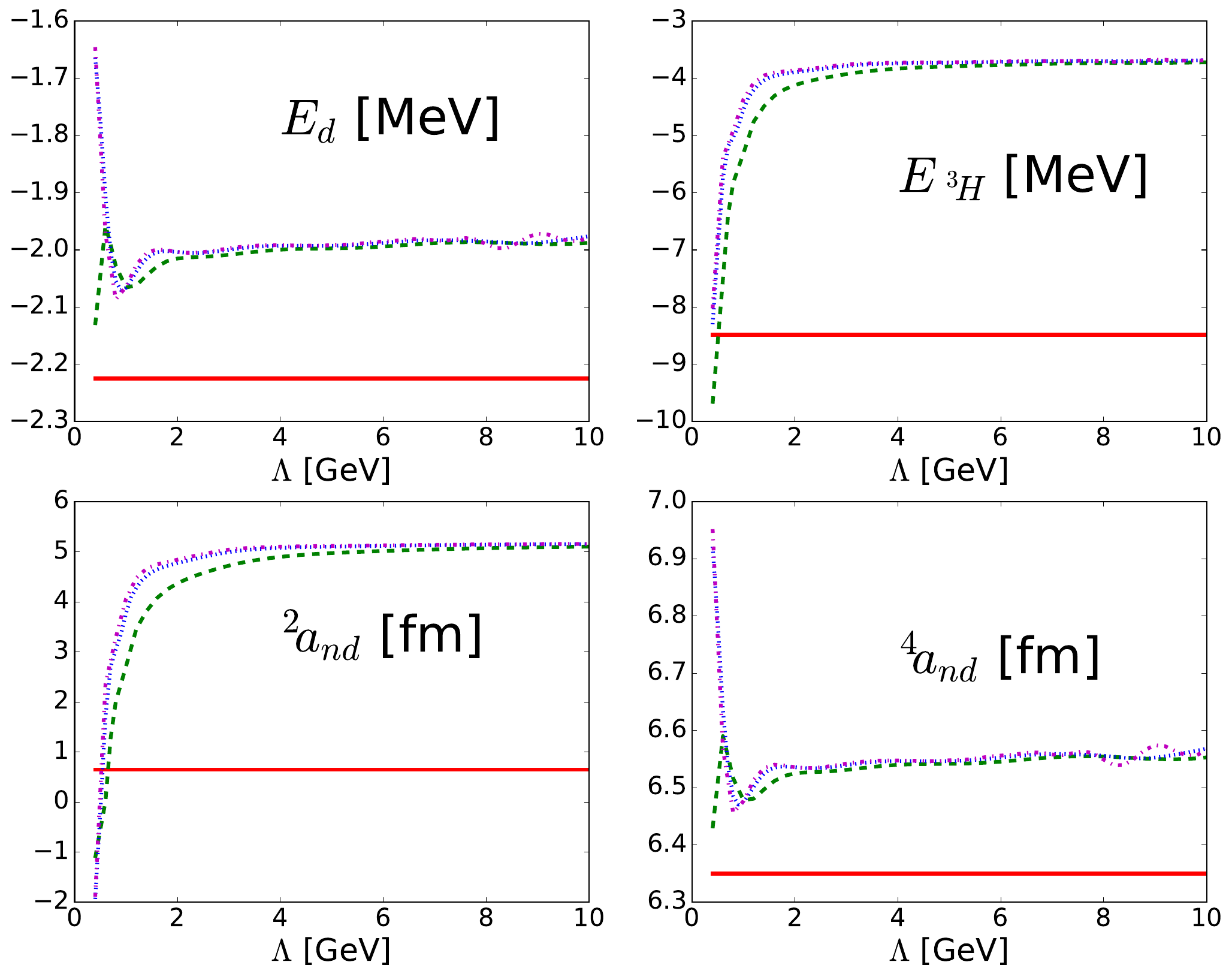}
\caption{(Color Online) Cutoff dependence of
         deuteron (top left) and triton (top right) energies, and
         $J=1/2$ (bottom left) and $J=3/2$ (bottom right)
         neutron-deuteron scattering lengths at LO,
         for different regulator choices, Eq. \eqref{regs}.
         Results are shown for $n=2$ (green dashed line),
         $n=4$ (blue dotted line), and $n=6$ (magenta dot-dashed line),
         in comparison with experiment (red solid line).
}
\label{fig:fig_3b_regulator}
\end{figure}

In addition, three-nucleon observables are insensitive to the form
of the regulator, Eq. \eqref{regs}, as shown
in Fig.~\ref{fig:fig_3b_regulator}.
Essentially the same converged values are achieved regardless of the
regulator choice, although as expected there is some regulator dependence
at low cutoff values. The
larger oscillations of $n=6$
results
can be attributed to artifacts from the numerical Fourier transformation.

\begin{figure}[tbp]
\centering
\includegraphics[width=0.8\linewidth]{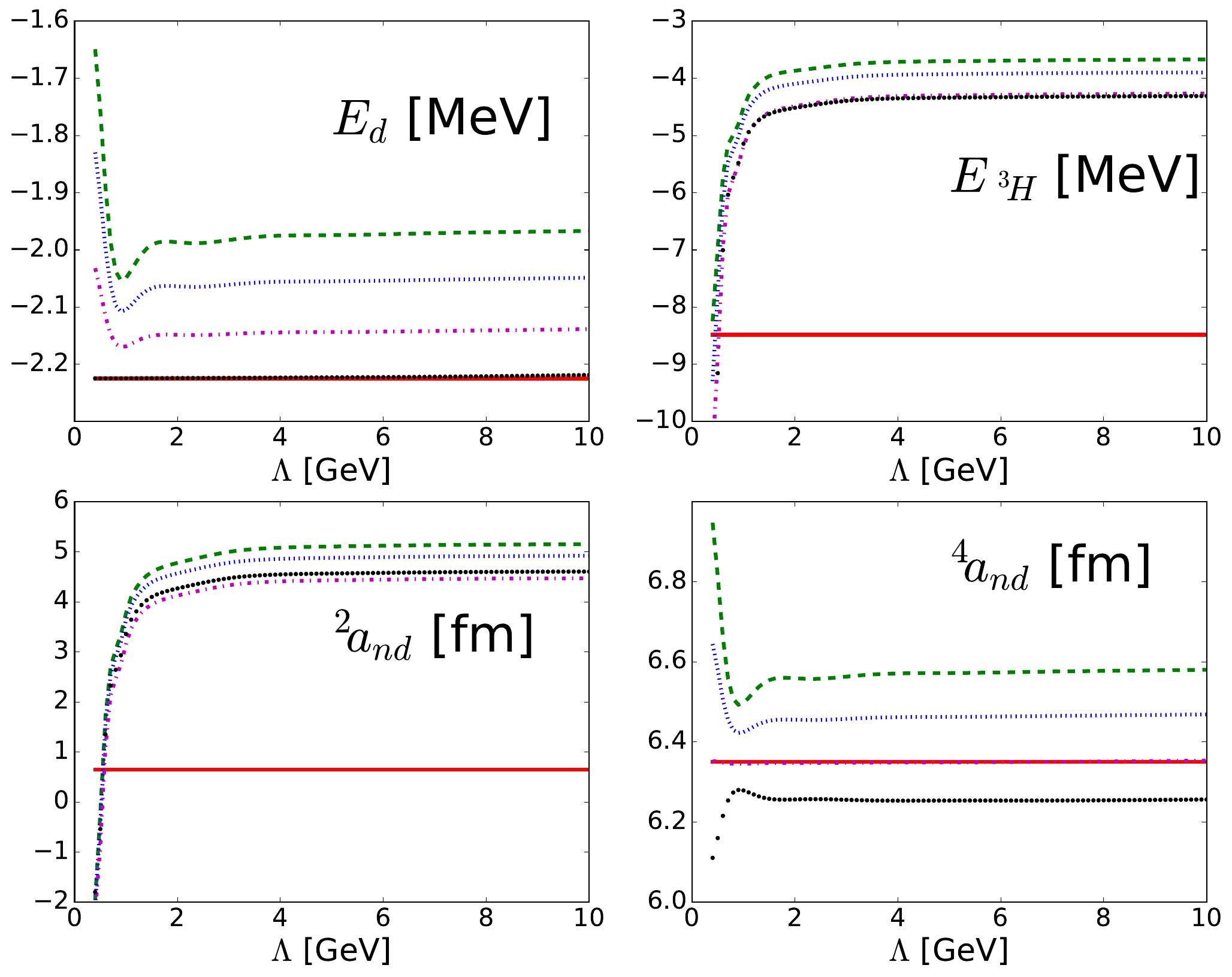}
\caption{(Color Online) Cutoff dependence of
the deuteron (top left) and triton (top right) energies,
and $J=1/2$ (bottom left) and $J=3/2$ (bottom right)
neutron-deuteron scattering lengths at LO with the $n=4$ regulator,
for different fitting procedures for the two-nucleon $S$-wave counterterms.
Results are shown for strategies
(I) (green dashed line),
(II) (blue dotted line),
(III) (magenta dot-dashed line),
and (IV) (black dots),
in comparison with experiment (red solid line).
}
\label{fig:fig_3b_fitenergy}
\end{figure}

We find that the triton
is underbound and
the $nd$ scattering length in the doublet channel is
rather large when compared to experiment.
This is consistent with
the underbinding of the deuteron
and the poor description
of $^1S_0$ phase shifts at LO.
Thus, we compare results from the different choices
of fitting procedure for $S$-wave counterterms
shown in Table~\ref{tbl:counter_term_fit}.
Figure~\ref{fig:fig_3b_fitenergy} shows that cutoff independence
is achieved for the three-nucleon system in all cases,
but with
some dependence on the counterterm fitting method.
This dependence is comparable to the variation coming from cutoff values
$\Lambda \simge 1$ GeV.
The deuteron binding energy and the $nd$ quartet
scattering length are closely correlated with the $^3S_1$ counterterm.
As the
latter changes such as to provide more attraction,
the $nd$ quartet scattering length decreases
and crosses the experimental value.
The triton binding energy and
the $nd$ doublet scattering length
also approach experimental values, but remain quite far.

\begin{figure}[tbp]
\centering
\includegraphics[width=1\linewidth]{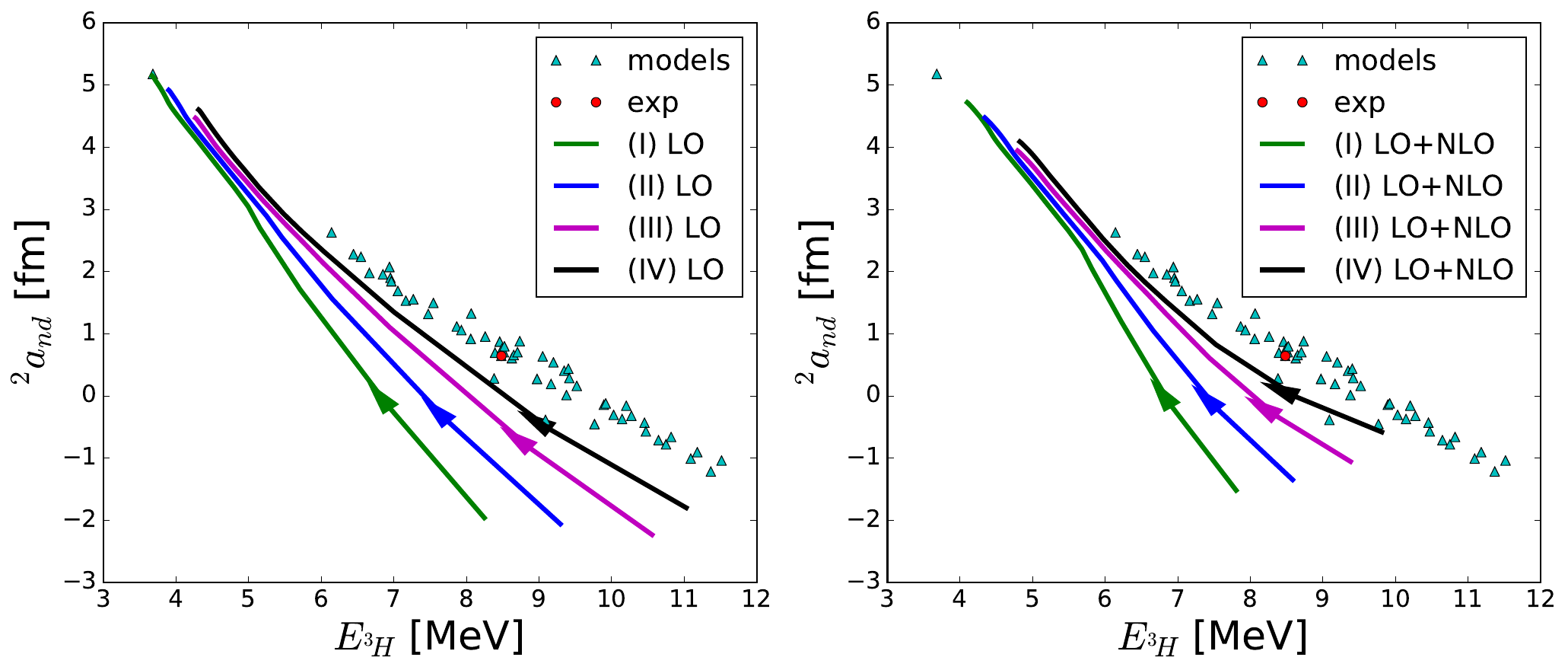}
\caption{(Color Online)
Correlation between triton binding energy and $nd$ doublet scattering length.
at leading order (left) and next-to-leading order (right).
Different lines correspond to cutoff variation 
from the various fitting strategies,
the arrow indicating the direction of increasing cutoff.
Phenomenological model results (triangles) and the empirical value (circle)
are also shown. 
}
\label{fig:fig_3b_NLO_phillips}
\end{figure}

We observe in our results a correlation between 
the triton binding energy and the $nd$ doublet scattering length
known as the Phillips line \cite{Phillips:1968zze}.
It has been observed that different models tend to align in 
the plane generated by all values of the triton binding energy and 
the $nd$ doublet scattering length.
In our case,
experimental values for these quantities
are obtained at low cutoff values (less than 1 GeV),
but are overshot by converged values.
In contrast, the deuteron energy and the $nd$ quartet scattering length
initially approach the experimental values but never reach them.
Our unconverged results can be seen as models that should 
produce a ``Phillips line'' as the cutoff is varied.
The left panel of Fig.~\ref{fig:fig_3b_NLO_phillips} shows
this correlation at LO, together with model calculations. 
The arrow in the graph indicates the direction of
increasing cutoff. Note that only the end point of each line
is our final result in the cutoff-independent limit. 
The spread of lines around these points gives a lower bound on the
expected contribution from the next order.
It is understandable that 
strategy (IV) is closest to the 
phenomenological Phillips line,
since phenomenological models are usually made to reproduce 
the two-nucleon effective-range parameters,
which determines the slope of the line.

\begin{figure}[tbp]
\centering
\includegraphics[width=0.8\linewidth]{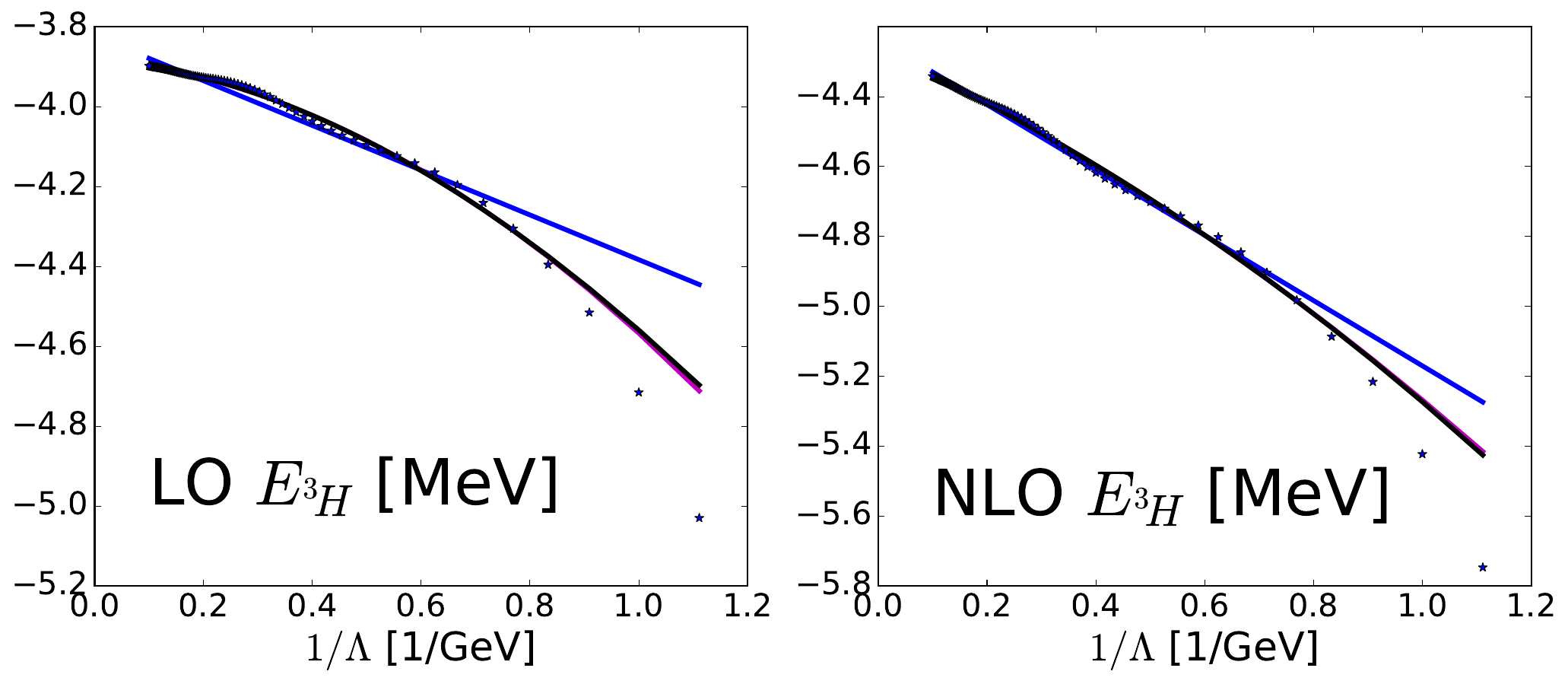}
\caption{(Color Online) Cutoff dependence of the triton binding energy
      at LO (left) and NLO (right) 
      with the $n=4$ regulator and fitting procedure (II).
      The blue, magenta, and black solid lines are, respectively, 
      two-, three-, and four-parameter fits with 
      Eqs. \eqref{tritonbindingfit:pol} and \eqref{tritonbindingfit:pol1} 
      to results for $\Lambda \geq 1.2$ GeV.
}
\label{fig:fig_3b_fits:triton}
\end{figure}

\begin{figure}[tbp]
\centering
\includegraphics[width=0.8\linewidth]{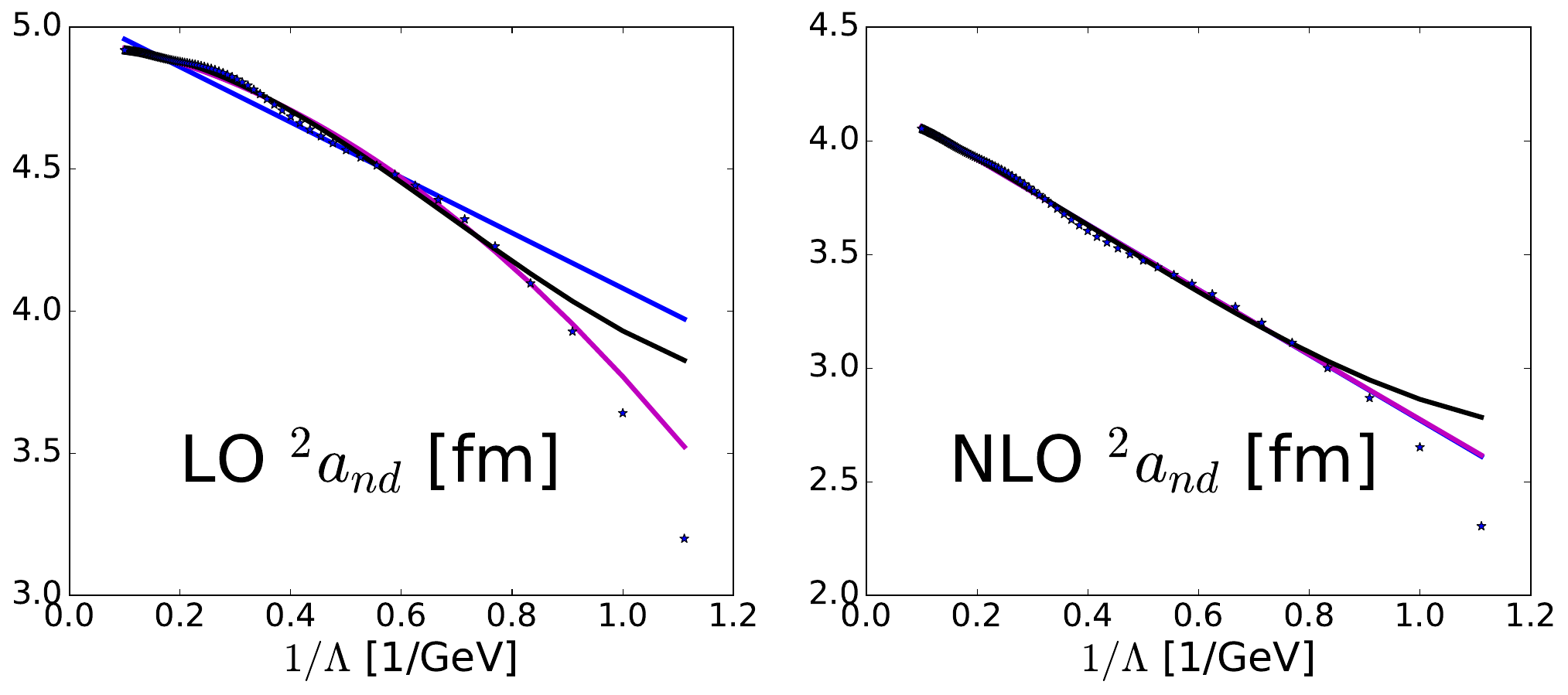}
\caption{(Color Online) Cutoff dependence of the $nd$ 
      doublet scattering length
      at LO (left) and NLO (right) 
      with the $n=4$ regulator and fitting procedure (II).
      The blue, magenta, and black solid lines are, respectively, 
      two-, three-, and four-parameter fits with 
      Eqs. \eqref{doubletscattlengthfit:pol} and 
      \eqref{doubletscattlengthfit:pol1} 
      to results for $\Lambda \geq 1.2$ GeV.
}
\label{fig:fig_3b_fits:ndJ1}
\end{figure}

The cutoff dependence of three-body observables
suggests that, unlike the Pionless EFT case,
there is no need for a three-body force at LO.
This is consistent with Weinberg's
power counting where three-body forces appear at higher order.
We attempt to infer the order of the short-range three-nucleon force
from the residual cutoff dependence of three-nucleon observables
most sensitive to this force: those in the doublet channel,
where the exclusion principle does not forbid the
three nucleons to be close together.
To be definite, we consider the results for the $n=4$ regulator
in the range $\Lambda\ge 1.2$ GeV,
with counterterms fitted to the $T_L=5$ MeV PWA93 data.
{}From RGI, we expect inverse powers of the cutoff
at large cutoff values,
so we fit the triton energy with a power series
\begin{equation}
E_{^3{\rm H}}^{(0)}(\Lambda)=E_{^3{\rm H}}^{(0)}(\infty)
 \left[1+ \frac{p_{t1}^{(0)}}{\Lambda}
 + \left(\frac{p_{t2}^{(0)}}{\Lambda}\right)^2
 + \left(\frac{p_{t3}^{(0)}}{\Lambda}\right)^3
+\cdots\right],
\label{tritonbindingfit:pol}
\end{equation}
with parameters $E_{^3{\rm H}}^{(0)}(\infty)$ and $p_{t1,2,3}^{(0)}$.
Likewise, the doublet scattering length is fitted by
\begin{equation}
^2a_{nd}^{(0)}(\Lambda)=\, ^2a_{nd}^{(0)}(\infty)
\left[1-\frac{p_{d1}^{(0)}}{\Lambda}+\left(\frac{p_{d2}^{(0)}}{\Lambda}\right)^2
     +\left(\frac{p_{d3}^{(0)}}{\Lambda}\right)^3 +\cdots \right],
\label{doubletscattlengthfit:pol}
\end{equation}
with parameters $^2a_{nd}^{(0)}(\infty)$ and $p_{d1,2,3}^{(0)}$.
The fitting parameters are summarized in Tables \ref{tbl:triton:fit}
and \ref{tbl:ndJ1:fit}, and 
the corresponding fits are shown in 
the left panels of Figs.~\ref{fig:fig_3b_fits:triton}
and \ref{fig:fig_3b_fits:ndJ1}. 
Numbers, particularly at the highest power of a truncation
and in the four-parameter fit,
depend somewhat on the cutoff range,
regulator function, and fitting procedure.
However, qualitative conclusions do not change.
After they stabilize, the momenta 
$p_{t1}^{(0)}\sim p_{d1}^{(0)}\sim p_{s1}^{(0)}$,
which is consistent with the existence of NLO corrections
in the $^1S_0$ channel. 
The four-parameter fits are afflicted by large errors
in the higher parameters  $p_{t2,3}^{(0)}$ and $p_{d2,3}^{(0)}$,
probably due to the oscillations ---
a consequence is the weird low-cutoff behavior of the four-parameter fit
of the doublet scattering length.
The large errors of this fit suggest
that, again, higher-power fits would be unreliable.
For the three-parameter fit, the momenta $p_{t2}^{(0)}\sim p_{d2}^{(0)}$
are somewhat large, possibly indicating larger N$^2$LO corrections.

\begin{table}
\begin{tabular}{c|c|c|c|c|c|c|c}
  \multicolumn{4}{c|}{LO}  & \multicolumn{4}{c}{NLO}     
\\
\hline
       $E_{^3H}^{(0)}(\infty) $ &  $p_{t1}^{(0)}$  & $p_{t2}^{(0)} $ 
      & $p_{t3}^{(0)} $ 
      & $E_{^3H}^{(1)}(\infty) $ &  $p_{t1}^{(1)}$  & $p_{t2}^{(1)}$ 
      & $p_{t3}^{(1)}$  
\\       
\hline
      $-3.82$ & $146(3)$ & -  & - &
             $-4.24$ & $221(2)$ & -  & -   
\\
      $-3.88$ & $35.8(4.6)$ & $377(75)$  & - &
             $-4.27$ & $167(5)$ & $262(80)$  & - 
\\
      $-3.88$ & $29.6(14)$ & $399(194)$  & $243(310)$ &
              $-4.26$ & $171(16)$ & $239(208)$  & $208(325)$ 
\\   
\end{tabular}
\caption{Parameters of fits using Eqs. \eqref{tritonbindingfit:pol} 
        and \eqref{tritonbindingfit:pol1} 
        to the leading-order and next-to-leading-order results 
        for the triton binding energy 
        from fitting procedure (II) for cutoff values $\Lambda\geq 1.2$ GeV.
        The parameters $E^{(m)}_{^3H}(\infty)$ and $p_{t1,2,3}^{(m)}$ ($m=0,1$)
        are in MeV.
        Numbers in parentheses represent fitting errors.
}
\label{tbl:triton:fit}
\end{table}

\begin{table}
\begin{tabular}{c|c|c|c|c|c|c|c}
    \multicolumn{4}{c|}{LO}  & \multicolumn{4}{c}{NLO}     
\\
\hline 
     ${}^2a_{nd}^{(0)}(\infty) $ &  $p_{d1}^{(0)}$  & $p_{d2}^{(0)} $ & $p_{d3}^{(0)} $
    & ${}^2a_{nd}^{(1)}(\infty) $ &  $p_{d1}^{(1)}$  & $p_{d2}^{(1)}$ & $p_{d3}^{(1)} $
\\
\hline 
  $5.06$ & $193(4)$ & -  & -
          & $4.21$ & $341(2)$ & - &- 
\\
    $4.96$ & $53.0(6.6)$ & $433(92)$  & -
           &  $4.21$ & $343(6)$ & $52(93)$ & - 
\\
    $4.92$ & $47.0(16.6)$ & $693(216)$  & $615(332)$
           &  $4.18 $ & $280(19) $ & $430(232) $ & $531(349)$ 
\\
\end{tabular}
\caption{Parameters of fits using Eqs. \eqref{doubletscattlengthfit:pol} 
        and \eqref{doubletscattlengthfit:pol1}
        to the leading-order and next-to-leading-order results 
        for the $nd$ doublet scattering length 
        from fitting procedure (II) for cutoff values $\Lambda\geq 1.2$ GeV.
        The parameters $^2a_{nd}^{(m)}$ and $p_{d1,2,3}^{(m)}$ ($m=0,1$)
        are in fm and MeV, respectively.
        Numbers in parentheses represent fitting errors.
}
\label{tbl:ndJ1:fit}
\end{table}

In summary,
the weak cutoff dependence of three-body observables suggests that
there is no need for a three-body counterterm at LO
in the RGI scheme of Nogga {\it et al.}~\cite{Nogga:2005hy}.
This result is consistent with Weinberg's original power
counting. However, considering the importance of NLO
in the $^1S_0$ two-nucleon partial wave, we are led to consider
the corresponding effects in the three-nucleon system.
We first return to the two-nucleon system to 
quantify the NLO improvements there.

\subsection{Two-nucleon system at NLO}

As discussed earlier,
an NLO
correction in the $^1S_0$ channel is included to comply with
RGI.
Also, as seen in
LO results,
the deviation of the $^1S_0$ phase shift from PWA93 data
appears already at low energies and
cannot be cured without the NLO correction.

\begin{figure}[tbp]
\centering
\includegraphics[width=1\linewidth]{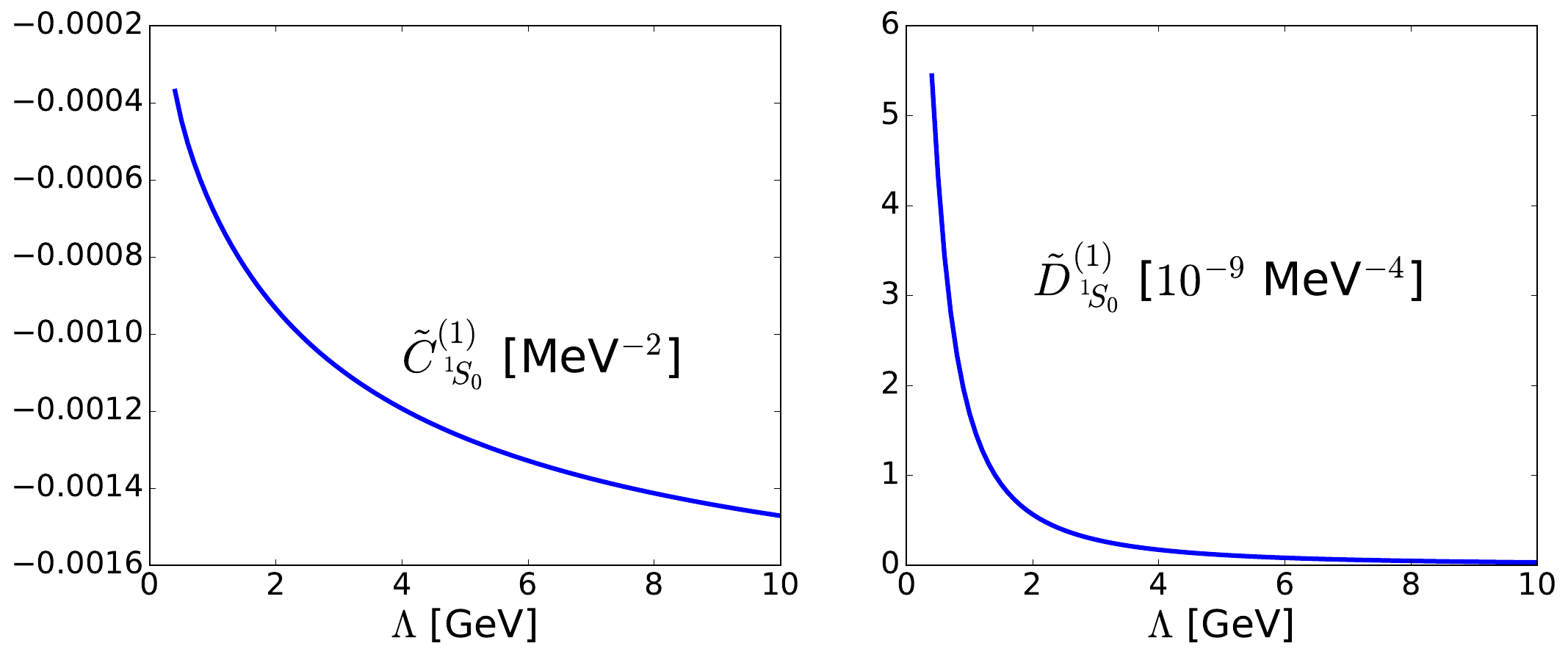}
\caption{(Color Online)
Cutoff dependence of the NLO counterterms
$\tilde{C}^{(1)}_{^1S_0}$ (left) and $\tilde{D}^{(1)}_{^1S_0}$ (right),
with the $n=4$ regulator.
Counterterms are fitted to the PWA93 phase shifts at 5 and 10 MeV.
}
\label{fig:fig_ct_nlo}
\end{figure}

For the determination of NLO counterterms, we compute
the NLO phase shift
from a DWBA calculation with the NLO potential.
We fit the counterterms to
reproduce phase shifts or effective-range parameters,
according to the three cases (I), (II) and (III) 
listed in Table \ref{tbl:counter_term_fit}.
As an example, Fig.~\ref{fig:fig_ct_nlo} shows the cutoff dependence of
the counterterms 
$\tilde{C}^{(1)}_{^1S_0}$ and $\tilde{D}^{(1)}_{^1S_0}$
for fitting strategy (II).
Similar dependence is found for other strategies.

\begin{figure}[tbp]
\centering
\includegraphics[width=1\linewidth]{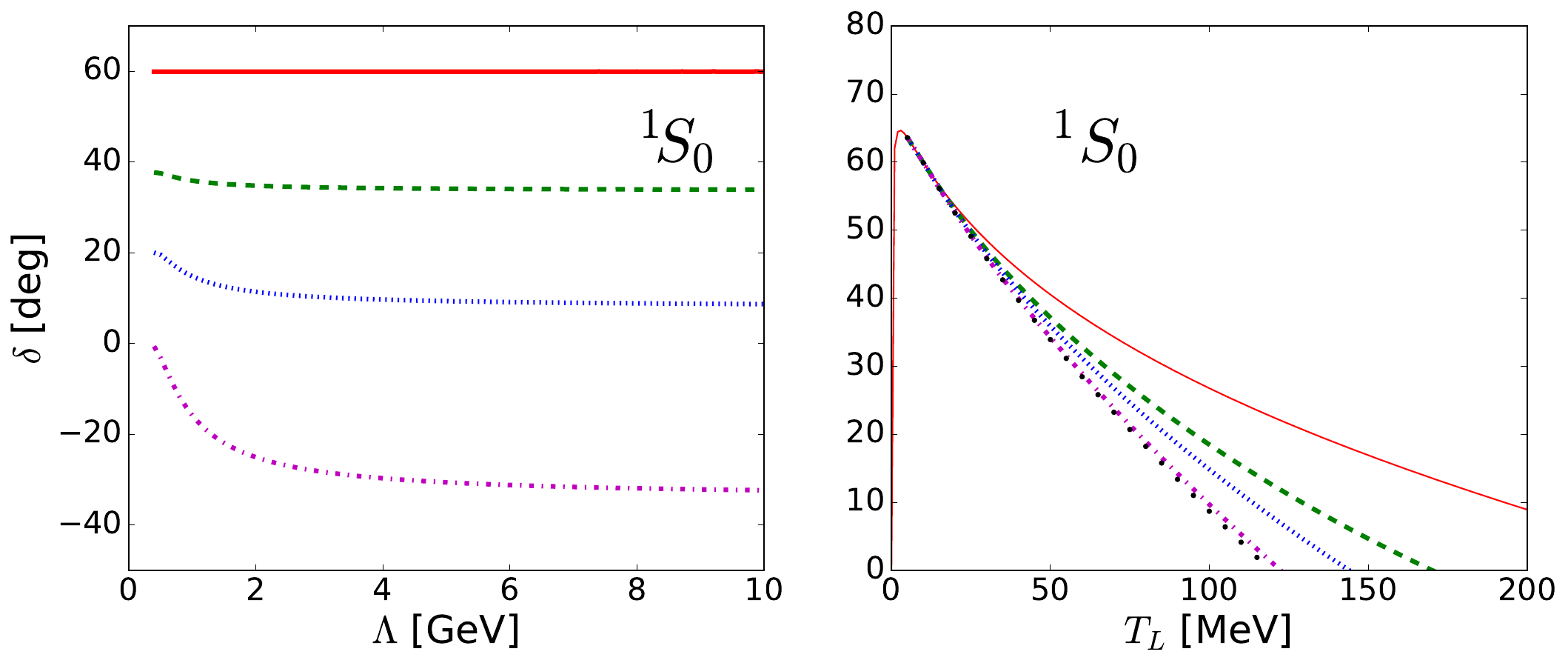}
\caption{(Color Online)
NLO $^1S_0$ phase shift with
the $n=4$ regulator, for fitting method (II).
Left panel: Cutoff dependence for various energies,
as in Fig.~\ref{fig:fig_2b_pwa_fitenergy_1s0} for LO.
Right panel: Lab energy dependence for various cutoffs,
as in Fig.~\ref{fig:fig_pwa_energy_singlet_all} for LO.
}
\label{fig:combo}
\end{figure}

In Fig. ~\ref{fig:combo} we display both
cutoff and energy dependence of the NLO $^1S_0$ phase shifts
for fitting procedure (II). 
We have also obtained similar curves for fitting procedures (I) and (III).
In comparison with Fig.~\ref{fig:fig_2b_pwa_fitenergy_1s0},
NLO results show decreased sensitivity
to the fitting method,
as desired in an EFT.
The low-energy phase shifts
are now found to be in good agreement with each other,
and they show much improved agreement with PWA93 data,
although there still remain deviations at larger energies.
Our result is similar to that of Long and Yang~\cite{Long:2012ve}
which used $T_L=5$ MeV and $25$ MeV for fitting the counterterms.
According to Ref. \cite{Long:2012ve}, a good
reproduction of PWA93 is achieved up to $T_L\simeq 100$ MeV,
once N$^2$LO (${\cal O}(Q^2/M_{hi}^2)$) corrections are included.

As for LO, we fit the NLO $^1S_0$ scattering length
results from the data fitting procedure (II) at 
$\Lambda \geq 1.2$ GeV
with a power series
\bea 
a_{s}^{(1)}(\Lambda)= a_{s}^{(1)}(\infty)
 \left[1+\frac{p_{s1}^{(1)}}{\Lambda}
 +\left(\frac{p_{s2}^{(1)}}{\Lambda}\right)^2
 +\left(\frac{p_{s3}^{(1)}}{\Lambda}\right)^3   
 +\cdots
 \right].
\label{1S0scattlengthfit:pol1}
\eea 
The fitting parameters $a_{s}^{(1)}(\infty)$ and $p_{s1,2,3}^{(1)}$ are 
also
summarized in Table \ref{tbl:1s0scattering:fit}
and the corresponding fits are shown in 
the right panel of Fig. \ref{fig:fit_2b_sc1s0:v2}.
The parameters are comparable to those at LO,
and
the asymptotic value, $a_{s}^{(1)}(\infty)$, is now closer to the empirical value.
A better description of the low-energy data at this order
can only be achieved with fitting strategy (IV).
The continuing existence of $\Lambda^{-1}$ dependence
indicates that the next correction in this channel appears at N$^2$LO, 
consistent with the expectation that 
two-pion exchange contributes at this order 
\cite{Ordonez:1992xp,Ordonez:1993tn,Ordonez:1995rz}.

In summary, NLO corrections significantly improve the description
of the two-nucleon $^1S_0$ phase shift,
and further improvement is expected one order higher.
We now turn to the effects of the NLO potential in three-nucleon observables.

\subsection{Three-nucleon system at NLO}

With the NLO
interaction so determined,
we compute the NLO corrections to three-nucleon observables.
Because the NLO
two-nucleon interaction acts only in the $^1S_0$ channel,  it does
not affect the deuteron binding energy and has little
effect on the $nd$ quartet scattering length, but it is significant
for the triton binding energy and the $nd$ doublet scattering
length.

\begin{figure}[tbp]
\centering
\includegraphics[width=1\linewidth]{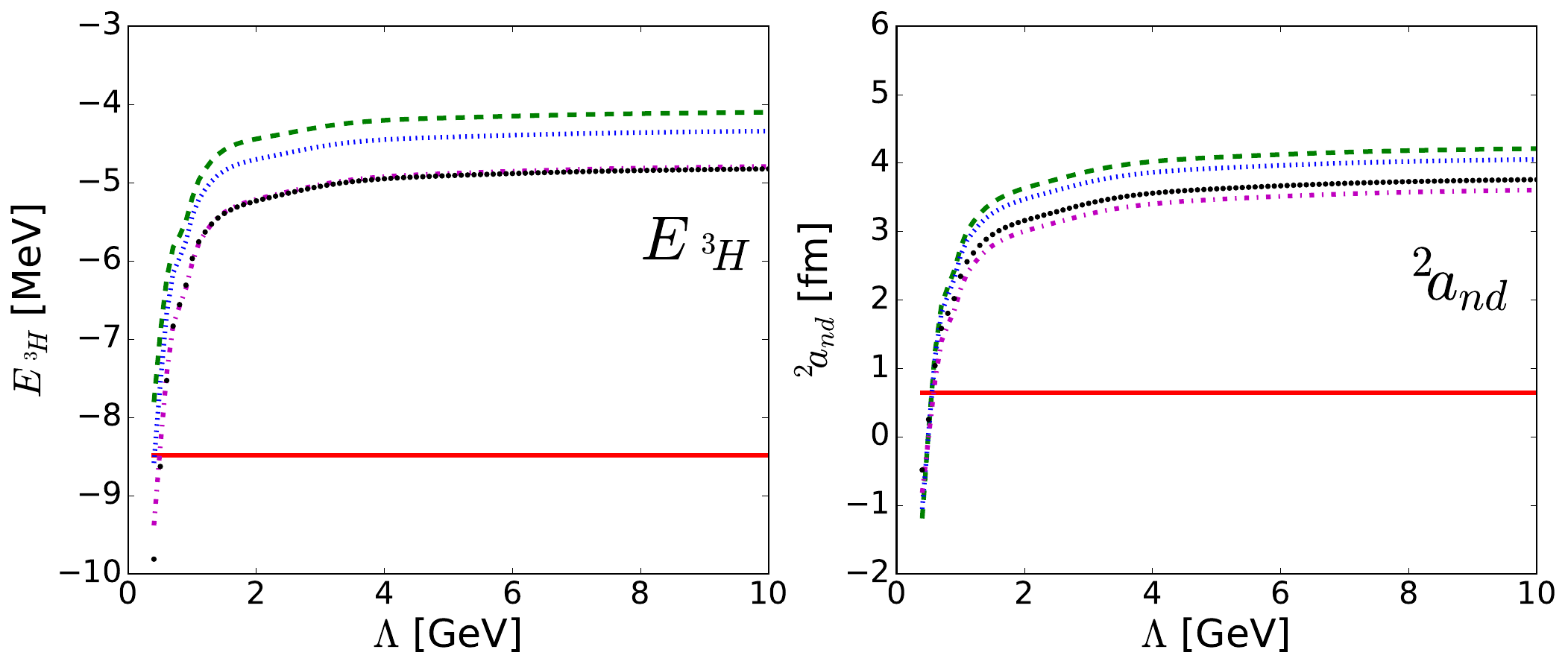}
\caption{(Color Online)
  Cutoff dependence of the triton energy (left)
  and the $J=1/2$ neutron-deuteron scattering length (right)
  at NLO with the $n=4$ regulator,
  for different fitting procedures for the two-nucleon $S$-wave counterterms.
  Curves as in Fig. \ref{fig:fig_3b_fitenergy}.
}
\label{fig:fig_3b_NLO_cut}
\end{figure}

Figure~\ref{fig:fig_3b_NLO_cut} shows the cutoff dependence of
these observables at NLO.
The graphs
include results with the different fitting procedures listed
in 
Table \ref{tbl:counter_term_fit}.
In all cases, cutoff independence is achieved at NLO,
and the residual cutoff dependence
at low cutoff values is reduced slightly.

The right panel of Fig.~\ref{fig:fig_3b_NLO_phillips} shows
the correlation between triton binding energy and
doublet scattering length at NLO, again together
with model calculations. 
The error estimated from the cutoff dependence reduces slightly from LO to NLO.
Also, the NLO result moves a bit towards the phenomenological
Phillips line.

To quantify the NLO residual cutoff dependence
we perform fits analogous to
Eqs. \eqref{tritonbindingfit:pol} and \eqref{doubletscattlengthfit:pol}:
\begin{equation}
E_{^3{\rm H}}^{(1)}(\Lambda)=E_{^3{\rm H}}^{(1)}(\infty)
 \left[1+ \frac{p_{t1}^{(1)}}{\Lambda}
 + \left(\frac{p_{t2}^{(1)}}{\Lambda}\right)^2
 + \left(\frac{p_{t3}^{(1)}}{\Lambda}\right)^3
+\cdots\right]
\label{tritonbindingfit:pol1}
\end{equation}
and
\begin{equation}
^2a_{nd}^{(1)}(\Lambda)=\, ^2a_{nd}^{(1)}(\infty)
\left[1-\frac{p_{d1}^{(1)}}{\Lambda}+\left(\frac{p_{d2}^{(1)}}{\Lambda}\right)^2
     +\left(\frac{p_{d3}^{(1)}}{\Lambda}\right)^3 +\cdots \right],
\label{doubletscattlengthfit:pol1}
\end{equation}
with asymptotic values $E_{^3{\rm H}}^{(1)}(\infty)$ and $^2a_{nd}^{(1)}(\infty)$,
and parameters $p_{t1,2,3}^{(1)}$ and $p_{d1,2,3}^{(1)}$.
We consider 
results for the $n=4$ regulator
in the range $\Lambda\ge 1.2$ GeV,
with counterterms fitted to the PWA93 data at $T_L=5$ MeV for LO and
$T_L=10$ MeV for NLO.
The fitting parameters are again summarized in Tables \ref{tbl:triton:fit}
and \ref{tbl:ndJ1:fit}, and 
the corresponding fits are shown in 
the right panels of Figs.~\ref{fig:fig_3b_fits:triton}
and \ref{fig:fig_3b_fits:ndJ1}. 

Convergence with the cutoff
implies that RGI is achieved up to NLO (${\cal O}(Q/M_{hi})$)
without the need of a short-range three-body force.
This is consistent with Weinberg's power counting.
The NLO asymptotic values $E_{^3{\rm H}}^{(1)}(\infty)$ and $^2a_{nd}^{(1)}(\infty)$
are closer to experiment than the LO asymptotic values 
$E_{^3{\rm H}}^{(0)}(\infty)$ and $^2a_{nd}^{(0)}(\infty)$.
Thus, NLO corrections reduce the difference between
theoretical and empirical values,
but they
are only
$\sim 1$ MeV for the triton binding energy
and $\sim 1$ fm for the $nd$ doublet scattering length.
The remaining discrepancy to experiment indicates
the presence of important corrections that we have not accounted for,
such as N$^2$LO (${\cal O}(Q^2/M_{hi}^2)$) corrections
or perhaps lower-order interactions not needed for RGI.
This is consistent with the increase in
the momenta $p_{t,d1}^{(1)}$ from the LO parameters $p_{t,d1}^{(0)}$.
As for 
LO, the four-parameter fits are plagued by large
errors and probably not much can be learned from them.

\section{Conclusion\label{sec:Summary}}

We have analyzed the cutoff dependence of
two- and three-nucleon observables at leading
and next-to-leading orders in the manifestly 
renormalization-group-invariant version
of Chiral EFT proposed in Ref. \cite{Nogga:2005hy}
and developed in Refs.
\cite{Valderrama:2009ei,Valderrama:2011mv,Long:2011qx,Long:2011xw,Long:2012ve}.
We have explored different regulator functions
and cutoff values up to 10 GeV,
as well as different fitting procedures.

At the two-nucleon level, our results agree with those in
Refs. \cite{Nogga:2005hy} and \cite{Long:2012ve}.
The two-nucleon interaction at LO produces results
that converge as the cutoff increases,
and are relatively insensitive to the regulator function
and fitting procedure.
The residual cutoff dependence at LO indicates the need for
an NLO (${\cal O}(Q/M_{hi})$) correction in the
spin-singlet $S$ wave.
Addition of such an interaction 
in perturbation theory improves the description
of phase shifts in this wave.
We thus constructed a Chiral EFT potential up to NLO that
produces a
two-nucleon amplitude consistent with
renormalization-group invariance.

With this potential, we have solved the Faddeev equation
to calculate
the triton binding energy and the two
neutron-deuteron scattering lengths.
At LO, the Faddeev equation is solved exactly (within
numerical precision), while at NLO first-order perturbation theory
is employed, as required by power counting.
Our LO result for the triton
binding energy again agrees with that of Ref. \cite{Nogga:2005hy}
for the same regulator function and 
cutoff range.
In addition, we significantly expanded the 
cutoff range
and studied the (in)sensitivity to the form
of the regulator function and choice of fitting input.
We also calculated the
neutron-deuteron scattering lengths for the first time.
We strengthen the conclusion of Ref. \cite{Nogga:2005hy}
that there is no renormalization need for a three-nucleon force
at LO.
We observe, again for the first time,
that three-nucleon observables display similar renormalization behavior
at NLO. Convergence with cutoff is achieved
for different regulator functions and fitting procedures.

Results in two- and three-nucleon systems show
that the modified power counting scheme of Nogga
{\it et al.} works well with respect to renormalization,
at least up to NLO.
The residual cutoff
dependence at NLO suggests that N$^2$LO corrections (${\cal O}(Q^2/M_{hi}^2)$)
are expected at both two- or three-nucleon levels.
No conflict has been seen with the higher order of
three-body forces expected on the basis of naive dimensional
analysis, as prescribed in Weinberg's original power counting.

We find that the three-nucleon observables we have calculated
are insensitive to two-nucleon waves
with angular momentum $l\ge 3$. This is in agreement with 
numerical \cite{Nogga:2005hy} and semi-analytical \cite{Birse:2005um} 
estimates of the importance of one-pion exchange
in the two-nucleon system.
Thus, our calculations are consistent with the naive expectation that
the two-nucleon waves with largest phase shifts at low energies
give the bulk of the contribution to few-body observables.
This expectation is captured in the power counting of Nogga
{\it et al.}, where only $l\le 2$ waves are treated non-perturbatively.
However, the transition in $l$ to subleading orders is not sharp,
and there is room for improvement in the treatment of various
two-nucleon waves \cite{PavonValderrama:2016lqn}.

Despite the apparent self-consistency of our calculation,
the triton 
is still considerably underbound at NLO, and the correlated
doublet $nd$ scattering length is much larger than experiment.
It should be remembered that Friar pointed
out \cite{Friar:1996zw} that the proper counting of factors of $4\pi$
implies that the dominant three-nucleon force in Chiral EFT
with explicit Delta isobars \cite{vanKolck:1994yi}
is also an NLO effect. We plan to return to this possibility
in a future publication.
Alternatively, the discrepancy seen here
might have an origin in the relatively large distances
that affect these quantities. In fact, they are very well described
in Pionless EFT already at LO \cite{Bedaque:1999ve},
indicating that cancelations must be present
in the higher-energy Chiral EFT.
Observables where cancelations occur are of course
not ideal to test the convergence of a theory: small
corrections at higher orders can generate relatively large changes
in observables
from the increased imbalance between partially canceling contributions.
This argument could be used as a rationale for the promotion of
a higher-order three-nucleon force,
as done very recently \cite{Kievsky:2016kzb}.
However, it is qualitatively different
than the need --- excluded here --- for promotion
mandated by renormalization,
when the very model independence of a calculation is at stake.
These issues show that much of
the optimal organization of Chiral EFT interactions
remains to be determined, a task
that requires the calculation
of a larger class of observables and the inclusion of higher orders.

\begin{acknowledgments}

We acknowledge useful discussions with J. Carbonell, Y. Kim,
and S. K\"onig.
YHS and UvK thank the Institute for Nuclear Theory
at the University of Washington
for its hospitality during the Program INT-16-1
``Nuclear Physics from Lattice QCD'', when part of this work was carried out.
This material is based upon work supported in part by
the Rare Isotope Science Project of the
Institute for Basic Science funded by Ministry of Science,
ICT and Future Planning and
National Research Foundation of Korea (2013M7A1A1075764),
by
the U.S. Department of Energy, Office of Science, Office of Nuclear Physics,
under award number DE-FG02-04ER41338,
and by
the European Union Research and Innovation program Horizon 2020
under grant No. 654002.

\end{acknowledgments}

\bibliography{EFT}

\end{document}